\documentclass[fleqn,usenatbib,useAMS]{mnras}

\usepackage{graphicx}
\usepackage{amsmath}
\usepackage{amssymb}
\usepackage{upgreek}
\usepackage{comment}
\usepackage{xspace}

\usepackage{times}
\usepackage{multicol}

\newcommand{\Mch}{$\text{M}_{\text{ch}}$\xspace}

\newcommand{\Msun}{\ensuremath{{\rm M}_{\sun}}\xspace}

\newcommand{\iso}[2]{\hbox{${}^{#1}{\text{#2}}$}\xspace}

\newcommand{\artis}{\textsc{artis}\xspace}
\newcommand{\cmfgen}{\textsc{cmfgen}\xspace}

\newcommand{\heatboostfour}{\textit{heatboost4}\xspace}
\newcommand{\heatboosteight}{\textit{heatboost8}\xspace}

\usepackage[T1]{fontenc}
\usepackage{ae,aecompl}

\usepackage{txfonts}

\title[]{Modelling the ionisation state of Type~Ia supernovae in the nebular-phase}

\author[Shingles et al.]{Luke J. Shingles$^{1,2}$\thanks{Email: luke.shingles@gmail.com},
Andreas Fl\"ors$^2$,
Stuart A. Sim$^1$,
Christine E. Collins$^2$,
\newauthor
Friedrich K. R\"opke$^{3,4}$,
Ivo R. Seitenzahl$^5$, and
Ken J. Shen$^6$
\\
$^1$Astrophysics Research Centre, School of Mathematics and Physics, Queen's University Belfast, Belfast BT7 1NN, Northern Ireland, UK\\
$^2$GSI Helmholtzzentrum f\"ur Schwerionenforschung, Planckstraße 1, 64291 Darmstadt, Germany\\
$^3$Zentrum f{\"u}r Astronomie der Universit{\"a}t Heidelberg, Institut f{\"u}r Theoretische Astrophysik, Philosophenweg 12, D-69120 Heidelberg, Germany\\
$^4$Heidelberger Institut f\"ur Theoretische Studien, Schloss-Wolfsbrunnenweg 35, D-69118, Heidelberg, Germany\\
$^5$School of Science, University of New South Wales, Australian Defence Force Academy, Canberra, ACT 2600, Australia\\
$^6$Department of Astronomy and Theoretical Astrophysics Center, University of California, Berkeley, CA 94720, USA
}

\date{Accepted 2022 March 30. Received 2022 March 29; in original form 2021 November 15}
\pubyear{2022}

\begin{document}
\label{firstpage}
\pagerange{\pageref{firstpage}--\pageref{lastpage}}
\maketitle
\begin{abstract}
	The nebular spectra of Type~Ia supernovae ($\gtrapprox$ 100 days after explosion) consist mainly of emission lines from singly- and doubly-ionised Fe-group nuclei. However, theoretical models for many scenarios predict that non-thermal ionisation leads to multiply-ionised species whose recombination photons ionise and deplete Fe$^{+}$, resulting in negligible [\ion{Fe}{II}] emission.
	We investigate a method to determine the collisional excitation conditions from [\ion{Fe}{II}] line ratios independently from the ionisation state and find that it cannot be applied to highly-ionised models due to the influence of recombination cascades on Fe$^{+}$ level populations. When the ionisation state is artificially lowered, the line ratios (and excitation conditions) are too similar to distinguish between explosion scenarios.
	We investigate changes to the treatment of non-thermal energy deposition as a way to reconcile over-ionised theoretical models with observations and find that a simple work function approximation provides closer agreement with the data for sub-\Mch models than a detailed Spencer-Fano treatment with widely used cross section data.
	To quantify the magnitude of additional heating processes that would be required to sufficiently reduce ionisation from fast leptons, we artificially boost the rate of energy loss to free electrons.
	We find that the equivalent of as much as an eight times increase to the plasma loss rate would be needed to reconcile the sub-\Mch model with observed spectra.
	Future studies could distinguish between reductions in the non-thermal ionisation rates and increased recombination rates, such as by clumping.
\end{abstract}

\begin{keywords}
radiative transfer -- supernovae: general -- white dwarfs -- line: formation -- atomic processes -- methods: numerical
\end{keywords}

\section{Introduction}
\label{sec:intro}

Type~Ia supernovae (SNe Ia) have been the subject of intense research for decades, which has often been motivated by their use as precise distance indicators for cosmology \citep{Perlmutter:1999fb,Riess:1998hp} and their role in synthesising much of the present-day abundance of Fe-peak nuclei \citep{Nomoto:1984jh,Matteucci:1986ux}.

While it has long been known that SNe Ia involve the thermonuclear explosion and subsequent unbinding of a carbon-oxygen white dwarf \citep[][]{Hoyle:1960bk,Bloom:2012he}, an understanding of the mechanism of explosion, the mass of the white dwarf, and the possible role of a binary companion remain open problems to be solved. Multi-band observations have placed deep upper limits on the density of the surrounding medium and disfavour most scenarios involving a non-degenerate companion \citep{Hancock:2011kf,PerezTorres:2014cb}. Several other scenarios remain, and thanks to theoretical developments in the last decade, we now have a variety of 3D hydrodynamic explosion models available \citep{Ropke:2018gi}. These explosion models can be post-processed with detailed nucleosynthesis calculations to obtain a snapshot of the nuclear densities seconds after explosion \citep{Seitenzahl:2017gu}. The simple time evolution of nuclear densities described by radioactive decays and homologous expansion can then be followed by radiative transfer calculations to produce synthetic light curves and spectra for comparison with observations.

\subsection{SNe Ia Nebular Spectra}
\label{sec:nebular_spectra}
Within seconds of the thermonuclear explosion of a white dwarf, ejected material expands rapidly with velocities of several 10$^3$\,km\,s$^{-1}$, causing the optical depth of the ejecta layers to continuously decrease. Roughly half a year after the explosion, the optical depth in the continuum and in many lines \citep[though there are exceptions, see e.g.,][]{Fransson:2015ct,Wilk:2020kb} have become so low that most optical and near-infrared photons can freely escape from the ejecta with few interactions. This phase is commonly referred to as the `nebular phase'.

\begin{figure*}
  \begin{center}
    \includegraphics[width=1\textwidth]{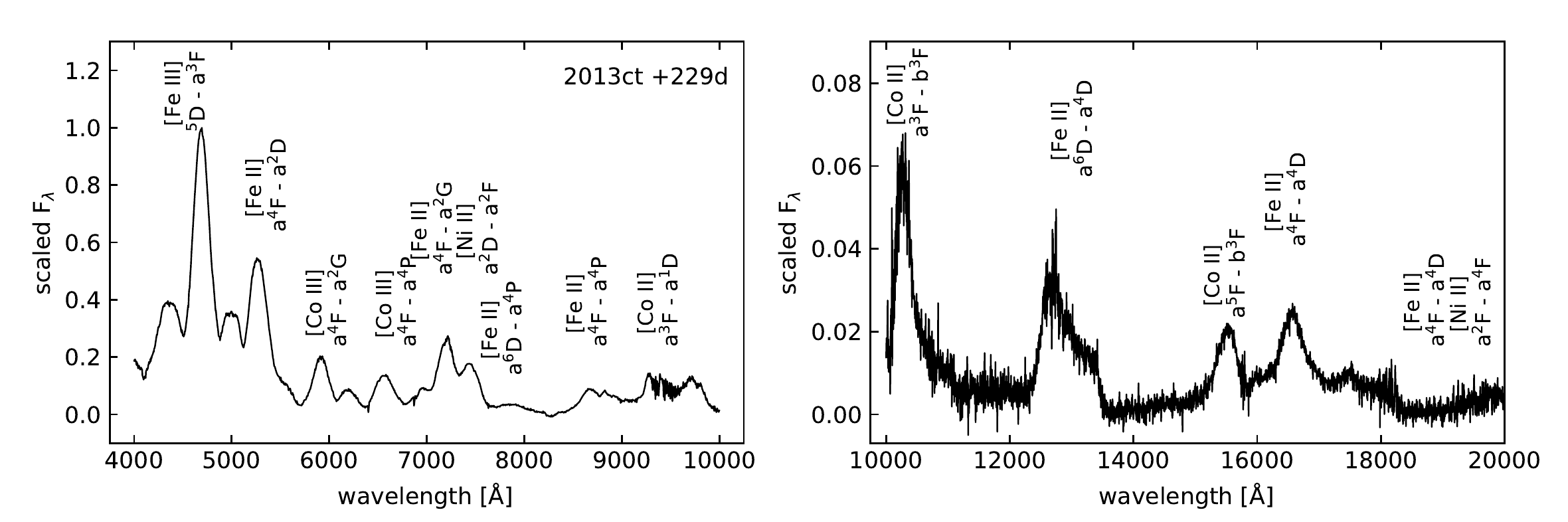}
   \end{center}
   \caption{Example spectrum of the normal SN\,2013ct in the nebular phase \citep{Maguire:2016jt}. Shown are the optical (left panel) and NIR (right panel) spectra as observed with X-Shooter at the Very Large Telescope (VLT) at time 229 days after B-band maximum. The multiplets with the strongest contribution to the observed features are labelled. We point out that the line fluxes are generally much weaker in the infrared than the optical region.}
    \label{fig:example_spectrum}
\end{figure*}

During the transition into the nebular phase, the observed spectrum changes from one that is dominated by continuum emission and the superposition of absorption/scattering lines on the continuum to an almost purely line-emission dominated spectrum (see the example in \autoref{fig:example_spectrum}). The emission lines originate from regions within the ejecta that are still being heated at these late phases by radioactive decays and thus contain \iso{56}{Co}. The outer layers, which are rich in intermediate mass elements or unburnt material, contribute less to the emission, but do produce features of S, Ca, and Ar.

The emission features in nebular spectra are broadened by the expansion velocity of the iron core (maximum extent $\sim 7,000$ to $\sim 12,000$\,km\,s$^{-1}$) from which the corresponding lines originate. 
Emission features have been observed and studied both in the optical and the near-IR (NIR, 3000\,\AA\,to 2.5\,$\upmu$m).

Singly- and doubly-ionised Fe is responsible for the majority of emission features in the optical and NIR (\autoref{fig:example_spectrum}). As the stable decay product of \iso{56}{Ni} decaying to \iso{56}{Co} and, finally, to \iso{56}{Fe}, Fe is readily abundant in the nebular phase. Its singly- and doubly-ionised ions have a large number of transitions that can be easily excited at the conditions (temperature and electron density) present during the nebular phase. In addition to Fe, Co and Ni are also seen in emission during the nebular phase.

The signature feature in the nebular phase spectrum of an SN Ia is located around 4,700\,\AA. The feature is created by emission from the $3d^6$\,$^5$D--$^3$F multiplet of [\ion{Fe}{III}] \citep{Axelrod:1980vk, Kuchner:1994fx}, with a likely contribution from a multitude of weak lines of [\ion{Fe}{II}]. The strength of the 4700\,\AA\,[\ion{Fe}{III}] feature with respect to a [\ion{Co}{III}] feature near 5900\,\AA, originating from the a$^4$F--a$^2$G multiplet, has been used as evidence for radioactive decay of \iso{56}{Co} to \iso{56}{Fe} being the power source of SNe Ia \citep{Kuchner:1994fx, Childress:2015hz,Dessart:2014jq}. The feature near 7,200\,\AA\,is a blend of mostly [\ion{Fe}{II}] and [\ion{Ni}{II}], with an additional contribution at its blue end from [\ion{Co}{III}] a$^4$F -- a$^4$P. A [\ion{Ca}{II}] contribution of $\lambda\lambda$7291, 7324 to this feature was seen in some radiative transfer models but was excluded by observed data \citep{Flors:2020ek}.

In the NIR, the observed emission originates mostly from [\ion{Fe}{II}], [\ion{Co}{II}] and [\ion{Ni}{II}]. It has been shown that the emission of [\ion{Co}{II}] relative to that of [\ion{Fe}{II}] declines as is expected from radioactive decay of \iso{56}{Ni} \citep[][]{Spyromilio:2004ho,Flors:2018cp,Flors:2020ek}. The relative strength of the [\ion{Ni}{II}] features in the optical and NIR is given just by the transition rates of the involved lines \citep[see][]{Flors:2020ek}.

\subsection{Theoretical challenges}

One of the main candidates for Type~Ia progenitors is a white dwarf with a mass well below the Chandrasekhar-mass limit, known as the sub-\Mch channel. The burning of a sub-\Mch white dwarf must proceed supersonically (by detonation) in order to synthesise the required abundances of intermediate-mass and Fe-peak nuclei \citep{Shigeyama:1992bq,Sim:2010hl,Shen:2018ed,Gronow:2020gc}. This channel is often contrasted with the near-\Mch (sometimes shortened to \Mch) channel, in which a white dwarf accretes material until its mass approaches the Chandrasekhar-limit and carbon fusion is ignited.

Among promising evidence for sub-\Mch progenitors, \citet{Flors:2020ek} use a semi-empirical method to work backwards from time-series spectra and infer Ni/Fe ratios that favour sub-\Mch over near-\Mch explosion models. However, modern radiative transfer studies of the same or similar, sub-\Mch and near-\Mch explosion models do not clearly favour sub-\Mch channel on the basis of predicted nebular spectra. It is therefore a major challenge for theorists to reconcile the opposite conclusions arrived at by the different approaches.

The semi-empirical approaches tend to focus on emission line ratios, such as between optical and NIR lines of [\ion{Fe}{II}], or between an [\ion{Fe}{II}] line and a [\ion{Ni}{II}] line. These quantities are chosen since the unknown ionisation state either cancels out \citep[for two lines of the same ion, e.g.,][]{Flors:2020ek} or is assumed to be the same for different Fe-group elements, Fe, Co, and Ni, \citep[as in ][]{Maguire:2018hp,Flors:2018cp,Flors:2020ek}. These assumptions are supported by similar ionisation energies of the relevant ionisation states (but see \citealt{Blondin:2021ur} and our \autoref{fig:submch-estimators} for differences between Ni$^{+}$/Ni and Fe$^{+}$/Fe) and spatially co-located Fe-group material, as is evident from the widths and shifts of the lines. These line ratio analyses favour sub-\Mch models mainly from abundance arguments: i.e., the relative strength of [\ion{Fe}{II}] to [\ion{Ni}{II}], and that is the main basis for their conclusions.

The forward-modelling approach, on the other hand, attempts to calculate a full spectrum from the conditions in the hydrodynamic explosion model, and that depends on the accurately modelling the full plasma state, such as the temperatures, ionisation and excitation conditions.

With the forward approach, the sub-\Mch models produce spectra that are clearly discrepant with the data. For example, \citet{Wilk:2018ki} find that their model spectra exhibit [\ion{Fe}{III}] $\lambda$4658 features that are too strong compared to other optical/NIR features, such as [\ion{Fe}{II}] lines. The main issue is that the relative strength of different ionisation states (e.g. [\ion{Fe}{III}] / [\ion{Fe}{II}]) is wrong. This issue is more problematic for the sub-\Mch models because the high \iso{56}{Ni} abundance throughout the core means that they are relatively highly ionised throughout the Fe-rich regions.

The nebular radiative transfer simulations of \citet{Shingles:2020gy} for sub-\Mch models are also discrepant with observations in a similar sense to \citet{Wilk:2018ki}, with relative weakness of [\ion{Fe}{II}] and [\ion{Ni}{II}] emission in the synthetic spectra.

To reduce the ionisation state of sub-\Mch models, \citet{Wilk:2020kb} examine the use of a clumping factor. This increases the local density, and hence the rates of collisional processes, such as recombination. Clumping can reduce the [\ion{Fe}{III}]/[\ion{Fe}{II}] ratio toward the level of observations. However, the low \iso{58}{Ni} abundance in their sub-\Mch model prevents a 1.939 $\upmu$m feature from forming, while for their near-\Mch-models, the [\ion{Ca}{II}] $\lambda\lambda$7291, 7324 feature becomes stronger than even the [\ion{Fe}{II}] $\lambda7155$, in conflict with observations.

The clumping result leads us to ask whether there could be any changes to the radiative transfer physics that would reduce the ionisation of the Fe-group in a targeted way without causing additional discrepancies. The fact that the ionisation and excitation states of distinct elements respond differently to clumping offers the hope of narrowing down possible causes of too-weak [\ion{Fe}{II}] emission by preferring candidates that improve the fit of spectral features from several different elements simultaneously. We justify this with the assumption that our models are approximately correct, and therefore the simplest solution to the discrepancy is more likely to be correct than a combination of several separate modelling problems.

\citet{Blondin:2021ur} experiment with manually adjusting the Ni$^{2+}$/Ni$^{{+}}$ ratio to match observed [\ion{Ni}{II}] features, either by boosting the ratio for a near-\Mch model or reducing the ionisation state of a sub-\Mch model. As possible physical explanations for the ionisation state discrepancy, \citet{Blondin:2021ur} consider clumping, as well as incorrect hydrodynamical model predictions of the densities and composition profiles.

The composition and densities of an explosion model affect the resulting ionisation state without changes to the radiation transfer method and ionisation treatment. \citet{Shingles:2020gy} find a good agreement with Type~Ia observations with the near-\Mch W7 spherically-symmetric fast-deflagration model, which produces strong emission by [\ion{Fe}{II}]. In this model, a core of stable $^{54,56}$Fe and $^{58}$Ni (i.e., no \iso{56}{Ni}) is heated non-locally, leading to the low ionisation state and moderate temperature needed for substantial [\ion{Fe}{II}] emission. However, while the fit to observations is interesting, the parameterised W7 model with a core devoid of radioactive material is unlike the predictions of sophisticated hydrodynamical simulations. In multi-dimensional delayed-detonation models, buoyancy mixing results in \iso{56}{Ni} being co-located with stable Fe-group material \citep{Seitenzahl:2013fz,Seitenzahl:2014iq}. In the case of a gravitationally-confined detonation, stable material is produced but at high velocities rather than in the centre \citep{Seitenzahl:2016cn}. In this work, we search for potential errors with the radiative transfer method that could cause the ionisation problem in the case that the most-sophisticated explosion models are approximately correct.

We explore the origin of the discrepancy between conclusions drawn from forward modelling of radiative transfer for a small set of multi-zone explosion models and those of semi-empirical analyses of observed spectra \citep[e.g.,][]{Flors:2020ek}. We aim to quantify the possibility that this discrepancy is mainly due to the sensitivity of synthetic spectra to the theoretically-predicted ionisation state and hence the rates of ionisation and recombination, while semi-empirical analyses concentrate on mass ratios that are inferred from singly-ionised line fluxes and statistical equilibrium solutions that neglect bound-free processes. We investigate the sensitivity of the observable line ratios and spectra to various treatments of the non-thermal deposition, and also test the validity of using one-zone approximations for modelling nebular emission lines, and the collisional-radiative approximation for calculating NLTE (Non-Local Thermodynamic Equilibrium) populations.

\subsection{Non-thermal ionisation}\label{sec:nonthermal}
In Type~Ia supernovae at late times, the thermalisation of decay products (mostly gamma-rays and $\beta^{{+}}$ from \iso{56}{Co} decays) is no longer efficient due to the low density and ionisation state of the ejecta. This leads to a population of fast leptons (electrons and positrons) with non-thermally distributed energies. Importantly, the impact ionisation by the non-thermal leptons overtakes photoionisation for most ions in the nebular phase. However, impact ionisation competes with two other processes by which the leptons lose energy -- exciting bound electrons and interacting with the slower thermally-distributed electrons. The thermal losses make up the vast majority (at least 85 per cent) of the non-thermal deposited energy.

\citet{Axelrod:1980vk} performed approximate calculations in the high-energy limit, estimating that about three per cent of the deposited energy contributes to ionisation in the conditions of nebular-phase Type~Ia supernovae. However, using currently-available atomic data with the detailed treatment of non-thermal deposition by \citet{Kozma:1992cy}, in which the Spencer-Fano equation \citep{Spencer:1954cb} is solved numerically, leads to a significantly larger fraction of the deposited energy causing ionisation. For the models of \citet{Shingles:2020gy}, the ionisation fraction can be as high as 15 per cent. The increased non-thermal ionisation rates, and consequent increase in photoionisation due to recombination photons from highly-charged (Fe$^{{3+}}$ and Fe$^{{4+}}$) ions, have a dramatic affect on the predicted nebular spectra of sub-\Mch models \citep{Shingles:2020gy}. Fe$^{+}$ is efficiently ionised to Fe$^{2+}$, which prevents formation of the strong [\ion{Fe}{II}] emission features that are observed in all nebular Type~Ia spectra.

The large difference in the fraction of deposition going to ionisation between the two methods exists despite similar ionisation cross sections (<10 per cent difference at 1.0 keV) for Fe between the \citet{Lotz:1967hq} formula used by \citet{Axelrod:1980vk} and the \citet{Arnaud:1992ft} formula fit (adopted by all known implementations of the \citet{Kozma:1992cy} method).

Although the \citet{Axelrod:1980vk} calculations are approximate, and do not consider further ionisation by secondary electrons, another source of the ionisation fraction difference is the use of a total bound-electron loss rate, for which the Bethe approximation is used. The total bound-electron loss rate used by \citet{Axelrod:1980vk} is greater than the summation of impact ionisation and non-thermal excitation cross sections and their transition energies that we, and other groups, use. The rate of energy loss to bound electrons can be up to a factor of two to four times higher using some form of the Bethe equation compared to a calculation of impact ionisation cross sections of Fe$^{{+}}$ and average energy loss of the primary electrons. See Appendix \ref{appendix:boundlossrate} for details.

The \citet{Axelrod:1980vk} method has no strict requirement that the sum of individual ionisation and excitation transitions are consistent with the total atomic loss rate. Any physical process that is implicitly counted in the total atomic loss rate, but not directly counted as an ionisation or excitation transition, contributes to the heating fraction. In contrast, a Spencer-Fano solution like \citet{Kozma:1992cy} does not take a total bound-electron loss rate as input, so every physical processes must be individually accounted for with a cross section in order to affect the electron degradation outcome.

The \citet{Kozma:1992cy} method has been successfully applied in the context of core-collapse supernova ejecta, which are mostly neutral with only a small fraction of singly-ionised material. The same implementation was also used by \citet{Jerkstrand:2015dz} to obtain spectra in reasonable agreement with nebular SN Ia observations from the W7 model. However, this was achieved without the contribution from ions that are triply-ionised and above. While for W7 this makes a subtle difference to the spectrum \citep[see figure 5 of ][]{Shingles:2020gy}, the consequences for sub-\Mch models are severe. The presence of recombination photons from Fe$^{3{+}}$ and Fe$^{4{+}}$ lead to a substantial increase in the photoionisation rate of Fe$^{{+}}$ \citep{Shingles:2020gy}.

A similar implementation of the Spencer-Fano solver in \cmfgen \citep{Li:2012dx} with the same ionisation cross sections \citep{Arnaud:1985wq,Arnaud:1992ft} results in over-ionised ejecta for sub-\Mch models \citep{Wilk:2018ki}. Directly comparing \cmfgen and \artis with identical explosion models finds good agreement for both the nebular spectra and ion fractions (Blondin et al. in prep).

The \citet{Kozma:1992cy} method is widely considered to be the best-available treatment of non-thermal deposition for Type~Ia models. The synthetic Type~Ia nebular spectra calculated with the \citet{Kozma:1992cy} method have never been directly compared against those that use the \citet{Axelrod:1980vk} approximation with all other modelling details held constant, and little investigation has been done into the uncertainties of using the \citet{Kozma:1992cy} method under Type Ia conditions with the widely-used electron-impact ionisation cross sections of \citet{Arnaud:1985wq} and \citet{Arnaud:1992ft}.

The simple \citet{Axelrod:1980vk} method is rarely used in recent Type~Ia models, but it is applied in recent models of kilonovae, which are the electromagnetic counterparts of Neutron-Star Mergers (NSM). For example, \citet{Hotokezaka:2021ui} present NSM models and use the \citet{Axelrod:1980vk} method, stating that a \citet{Kozma:1992cy} treatment would improve the accuracy of their results (presumably on the basis that the \citet{Kozma:1992cy} method is widely used in Type~Ia models).

The first kilonova model with a \citet{Kozma:1992cy} treatment of non-thermal ionisation is that of \citet{Pognan:2021ww}. However, in the absence of exerimental data, their applied impact ionisation cross sections are merely approximated. Alternative non-thermal ionisation treatments are not directly compared, and the validity of the calculations made using a Spencer-Fano approach with theoretical cross-sections is not tested.

Other NSM models in the literature that do not apply the \citet{Axelrod:1980vk} or \citet{Kozma:1992cy} methods depend on a total stopping power formula in order to model the thermalisation of fast electrons from $\beta$-decays \citep[e.g.,][]{Hotokezaka:2020jf}. The thermalisation efficiency results of \citet{Barnes:2016fu}, and studies that apply these analytic fits \citep[e.g.,][]{Kasen:2017kk,Wollaeger:2018gr} rely on a formula for the total energy loss rate \citep{Heitler:1954uu,Berger:1964vc,Chan:1993ic,Milne:1999gw}. Since the total stopping powers are either directly experimentally measured, or at least calibrated against stopping power measurements for other particles \citep[e.g., NIST ESTAR database][]{Estar:2017iu}, these quantities are expected to be reliably determined.

Future progress in producing accurate synthetic nebular spectra for kilonovae will require a detailed understanding of the non-thermal processes, and should be informed by studies into the accuracy of the \citet{Kozma:1992cy} method when implemented with the widely-used atomic data and approximations.

In this study, we directly compare the predicted nebular spectra for Type~Ia models that result from applying either the \citet{Axelrod:1980vk} limit, or the \citet{Kozma:1992cy} detailed Spencer-Fano solution. We investigate the consequences of a possible underestimation of the heating contribution from fast leptons in a Spencer-Fano solution by artificially boosting the plasma loss rate. This provides a single parameter that can transfer energy from the fast lepton population to the (already-dominant) thermal heating and reduce the non-thermal ionisation rates.

\section{Models and Method}

In this section, we describe the numerical methods and numerical test cases used to investigate the sensitivity of the nebular spectra and spectral line ratios to the non-thermal ionisation treatment, NLTE method, and the one-zone assumption used in semi-empirical method.

\subsection{Sub-\Mch detonation and W7 near-\Mch explosion models}
We use the \citet{Shen:2018ed} model of the detonation of a 1.0 \Msun white dwarf of Solar metallicity with an equal abundance of carbon and oxygen. This explosion model is very similar to the S0 model of \citet{Shingles:2020gy}, and therefore \artis radiative transfer calculations produce similar optical and near-infrared spectra as for S0 (notably lacking [\ion{Fe}{II}] emission).
For reference, we include results from the same radiative transfer simulation of the W7 explosion model \citep{Nomoto:1984jh,Iwamoto:1999jd} that was published by \citet{Shingles:2020gy}.

\begin{figure}
  \begin{center}
    \includegraphics[width=0.5\textwidth]{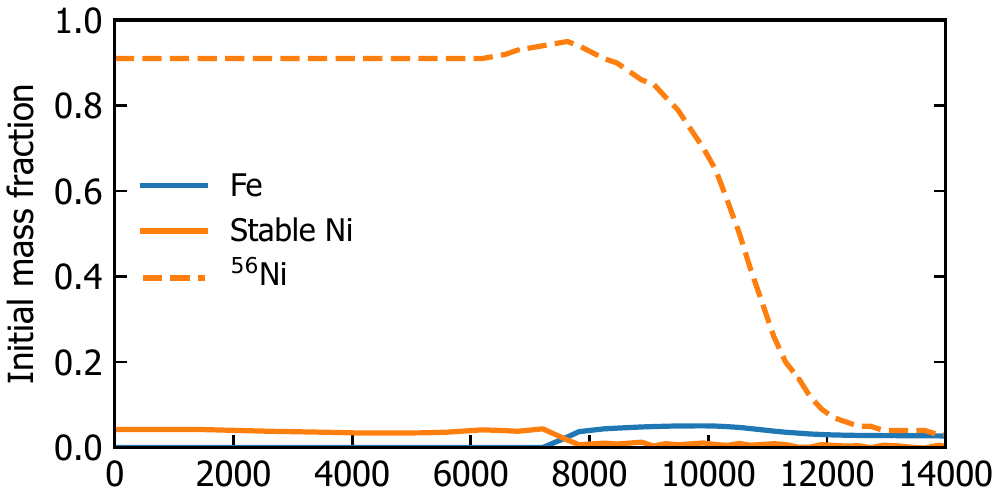}
    \includegraphics[width=0.5\textwidth]{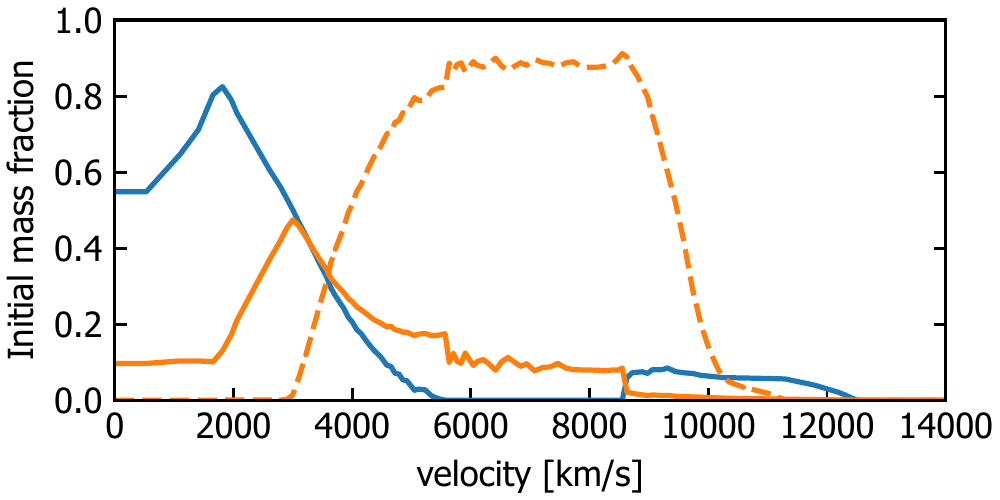}
  \end{center}
\caption{Composition profiles of the W7model 20 seconds after explosion (top) and the sub-\Mch model 10 seconds after explosion. The W7 model is that of \citet{Nomoto:1984jh} with the nucleosynthesis of \citet{Iwamoto:1999jd}. The sub-\Mch model is the 1.0 \Msun model Solar-metallicity 50\%-50\% C-O model of \citet{Shen:2018ed}.}
\label{fig:modelcompositions}
\end{figure}

\begin{table}
{
\centering
	\caption{Synthesised masses of \iso{56}{Ni}, \iso{54}{Fe}, and \iso{58}{Ni} from the W7 and sub-\Mch explosion models.}
	\label{tab:explosionmodels}
	\begin{tabular}{lccr}
		\hline
		& \multicolumn{3}{c}{Synthesised mass}\\
		Model & \iso{56}{Ni} & \iso{54}{Fe} & \iso{58}{Ni}\\
		 & $[\Msun]$ & $[10^{-2}\,\Msun]$ & $[10^{-2}\,\Msun]$\\
		\hline
		W7	& 0.59 & 9.5 & 11.0\\
		sub-\Mch		& 0.54 & 2.3 & 1.5\\
		\hline
	\end{tabular}\\}
\end{table}

\autoref{fig:modelcompositions} shows the post-explosion composition that is evolved by the \artis code according to radioactive decay and homologous expansion for use in the spectral synthesis calculation.

\subsection{Radiative transfer modelling}
We model radiative transfer in SNe~Ia using \artis\footnote{\href{https://github.com/artis-mcrt/artis/}{https://github.com/artis-mcrt/artis/}}, a Monte Carlo radiative transfer code that uses indivisible energy packets \citep{Lucy:2002hw}. The code has been described by \citet{Sim:2007is} and \citet{Kromer:2009hv} \citep[see also][for a description of the method]{Lucy:2005cx}. We use the latest version with NLTE enhancements described by \citet{Shingles:2020gy}, which include deposition by non-thermal electrons, NLTE ionisation balance and level populations, detailed bound-free photoionisation estimators \citep[using full packet trajectories as per equation 44 of][]{Lucy:2003bz} for all photoionisation transitions, and an NLTE binned radiation field model for excitation transitions.

\subsection{Non-thermal ionisation test cases}
As mentioned in Section \ref{sec:nonthermal}, the \citet{Kozma:1992cy} solution to the Spencer-Fano equation results in a lower heating fraction, and higher ionisation rates than the \citet{Axelrod:1980vk} high-energy limit approximation.

We directly compare the two methods with our `AxelrodNT' models, which replace the Spencer-Fano solution with the \citet{Axelrod:1980vk} work function approximation for non-thermal ionisation rates, and contribute 97 per cent of the energy deposited from gamma rays and positrons to heating.

We also experiment with sub-\Mch models (\heatboostfour and \heatboosteight) that have boost factors of four or eight applied to the plasma loss rate \citep{Schunk:1971fs,Kozma:1992cy} that accounts for Coulomb scattering of fast leptons by free (thermal) electrons. Increasing the assumed interaction strength between fast electrons and thermal electrons leaves less energy remaining for the collisional ionisation, shifting the Spencer-Fano solution in the direction toward the high (97 per cent) heating fraction of the \citet{Axelrod:1980vk} method. We stress that applying such a boost factor is not physically motivated but rather provides a convenient means to illustrate how changes in the deposition treatment affect the simulations.

\subsection{Parameterised model}\label{sec:parameterisedmodel}

The empirical analysis of \citet{Flors:2018cp,Flors:2020ek} makes use of a simplified model for the emission of the ejecta. In order to test assumptions of this approach, we aim to compare its predictions for key atomic level populations and line ratios to the \artis results.

The Flörs model does not explicitly solve for the plasma conditions of the ejecta (ionisation state, electron temperature and free electron density), but rather treats them as free parameters. By using the \artis plasma state solution as input for the simplified model, we will compare the results of the two NLTE population solutions. The \citet{Flors:2020ek} NLTE solution includes only bound-bound collisional and bound-bound downward radiative rates, i.e., in contrast to \artis, the Flörs model does not include radiative excitation, photoionisation or recombination transitions.

The energy source for excitation processes is characterised by a Boltzmann distribution of temperature $T$ and a free electron density $n_e$. In the simplified model, radiative transfer effects apart from self-absorption in the line of interest are ignored. Furthermore, non-thermal effects from the slowing-down of high-energy electrons, charge exchange reactions, and time-dependent terms in the NLTE rate equations are not included.
While sanity checks for the emitted radiation are included (e.g. the model cannot emit more radiation than is being produced through radioactive decay of \iso{56}{Ni}), the detailed energy deposition processes are not treated. As a result, the model is unable to provide the emitted radiation over all wavelengths, but only for certain lines in the optical and NIR.

To compute absolute level populations of the Fe-group ions, we use the ionisation balance from the \artis code, since the empirical approach does not involve a direct estimation of the ion population. Level populations within each ion, however, are computed with the code presented by \citet{Flors:2018cp,Flors:2020ek}.

To facilitate a comparison with the \artis code we changed the atomic data from the sources given in \citet{Flors:2020ek} to the CMFGEN database that is used by the \artis code. The same Fe$^{{+}}$ atomic data of \citet{Nahar:1994ft} are used in all cases. As a result, we can exclude differences in the atomic data as the reason for any difference in findings between this simplified model and the full radiative transfer simulations. Overall, the use of different atomic data sources only leads to subtle differences in the extracted quantities.

\section{Results}
We present the results from our radiative transfer models for the W7 and sub-\Mch explosions with several test cases for the non-thermal ionisation treatment. Throughout this section, we compare results with those obtained from the simplified model approach of Section \ref{sec:parameterisedmodel}, and, where appropriate, also with observations of the normal Type~Ia supernovae SN 2013ct.

In Section \ref{sec:modelstructure}, we present the internal conditions of the sub-\Mch model and changes in the \heatboostfour test case.
We then look at changes to emission spectra of individual ions in Section \ref{sec:ionspectra}.
In Section \ref{sec:spectra}, we compare the synthetic spectra of W7 and sub-\Mch models using various non-thermal ionisation treatments against observations.
In Section \ref{sec:lineratios}, we measure the ratios between the [\ion{Fe}{II}] $\lambda12570$ and $\lambda7155$, lines in our synthetic spectra for comparison to the semi-empirical trend line of \citet{Flors:2020ek}.
In Section \ref{sec:emittingconditions}, we check the accuracy of using line ratios to infer a single emitting plasma state by using the known emission conditions in our theoretical models.

\subsection{Structure of the sub-\Mch model}\label{sec:modelstructure}

\begin{figure*}
 \begin{center}\includegraphics[width=0.5\textwidth]{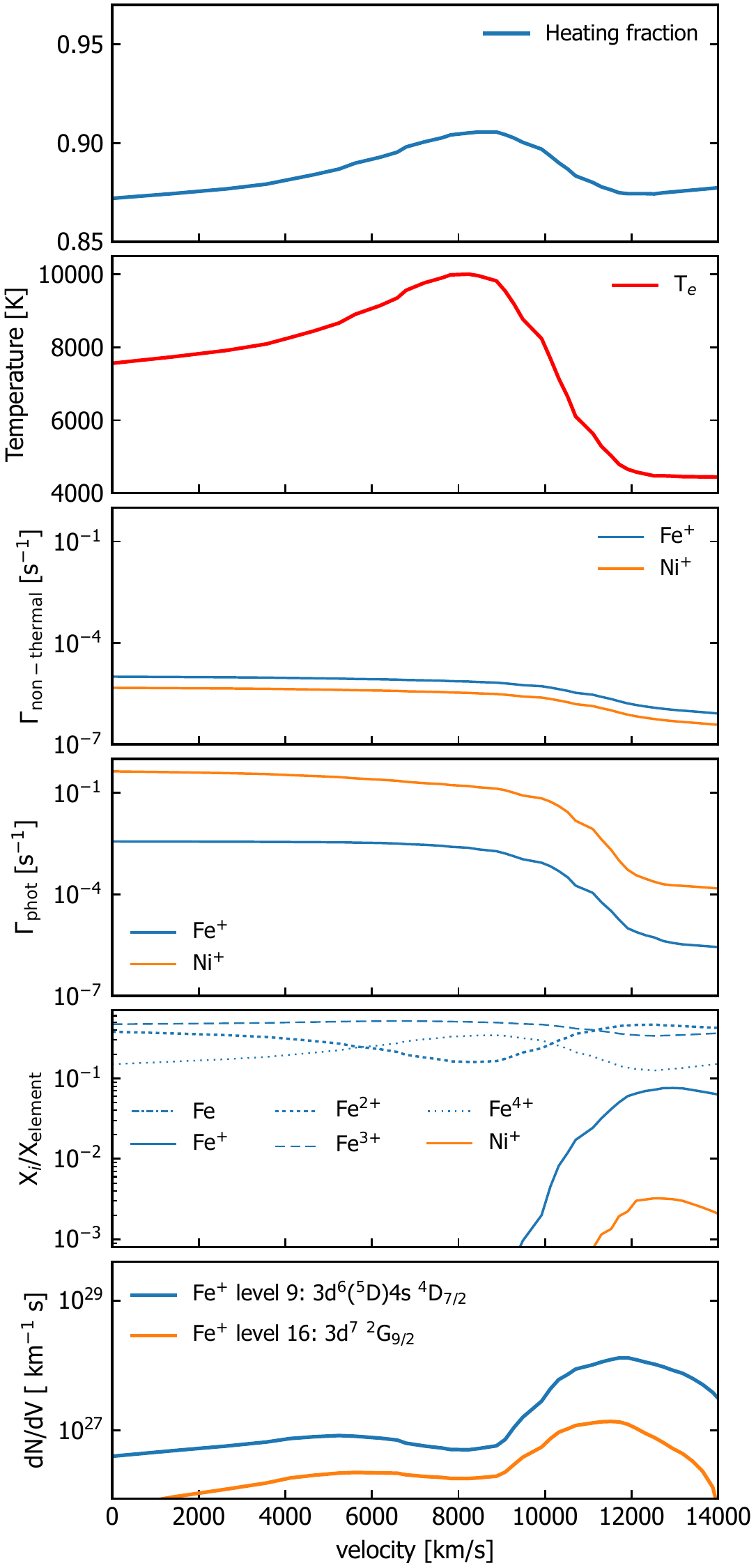}\includegraphics[width=0.5\textwidth]{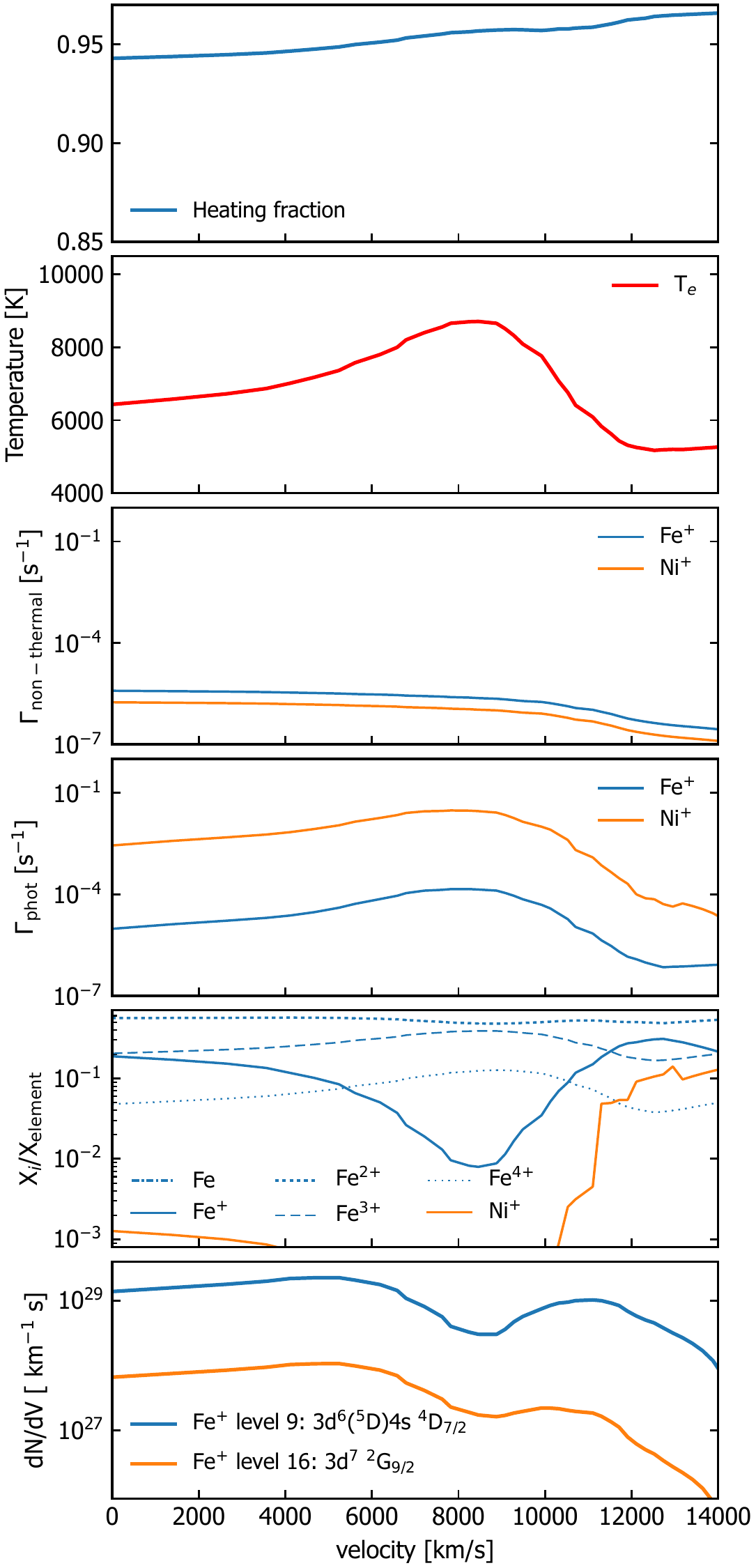}\end{center}
 \caption{Model structure of sub-\Mch detonation models at 247 days. The left panels are for the standard case, while the right is the \heatboostfour model with a factor of four boost in the electron loss rate of non-thermal particles. As a function of velocity in the model, the rows from top to bottom show: (1) the fraction of deposition energy contributed to thermal heating, (2) the electron temperature, (3) the non-thermal rate coefficient for singly-ionised Fe and Ni, photoionisation rate estimators for singly-ionised Fe and Ni, (4) the ion fractions of Fe for several ionisation states and the fraction of Ni in the form of Ni$^{{+}}$, and (5) the level population density per velocity interval for the Fe$^{{+}}$ upper levels of the [\ion{Fe}{II}] $\lambda12570$ (a$^4$D$_{7/2}$) and $\lambda7155$ (a$^2$G$_{9/2}$) lines. Bumpy features in the plot lines are due to Monte Carlo noise and shell boundaries. We note that there are differences between the Ni$^{+}$/Ni and Fe$^{+}$/Fe profiles. \label{fig:submch-estimators}}
\end{figure*}

As shown in \autoref{fig:submch-estimators}, the standard \artis calculation predicts that between 86 and 91 per cent of the energy deposited by non-thermal particles contributes to the heating of the thermal electrons throughout the model at a typical nebular phase (shown here for 247 days post explosion).
In its dense inner regions, the thermal electron temperature of the sub-\Mch model is between 7000 and 10000~K and
Fe is mainly ionised to the third and fourth stages (Fe$^{2{+}}$ and Fe$^{3{+}}$, see third panels in \autoref{fig:submch-estimators}). Some Fe$^{{+}}$ is present, but only in the low-density outer layers, which are significantly cooler (approximately 4000~K).

Comparison of the left and right panels in \autoref{fig:submch-estimators} shows that there are very important differences in the simulation if the electron-loss rate is boosted by a factor of four. First, as shown in the top panels, the fraction of energy that goes to heating increases, although only by $\sim 7$ per cent, which may seem relatively modest. The electron temperature (second row of \autoref{fig:submch-estimators}) also changes -- typically by about 2000~K, and typically the electron temperature drops despite the heating rate having increased. The reason for this drop in temperature can be understood from consideration of the ionisation fractions, as shown in the fifth row of \autoref{fig:submch-estimators}. With the boosted electron-loss function,
a substantial amount of Fe$^{{+}}$ is present close to the centre and decreasing to a minimum of approximately one per cent at around 8500 km/s. This significantly increases the cooling rate in most of the model, leading to an overall reduction in temperature for most zones.

The reason for the increase in Fe$^{{+}}$ abundance in the heatboost case with reduced non-thermal ionisation is not trivial and involves photoionisation. As found by \citet{Shingles:2020gy} for a similar sub-\Mch model S0, we also find that the photoionisation rate of Fe$^{{+}}$ far exceeds the non-thermal ionisation rate in our sub-\Mch model (see $\Gamma_{\rm phot}$ and $\Gamma_{\rm non-thermal}$ in \autoref{fig:submch-estimators}). For example, at 247 days in the shell at 10,000 km/s, the photoionisation rate of Fe$^{{+}}$ is more than one hundred times higher than its non-thermal ionisation rate. Therefore, decreasing only the non-thermal ionisation rate of Fe$^{{+}}$ would have a negligible effect on the Fe$^{{+}}$ fraction. Instead, a reduction in the non-thermal ionisation rates (by a factor of 2.9 for \heatboostfour and 4.9 for \heatboosteight) affects ionisation of Fe$^{{2+}}$ and Fe$^{{3+}}$ (which are not photoionisation-dominated) and decreases the abundances of higher ionisation stages. With fewer Fe$^{{3+}}$ and Fe$^{{4+}}$ ions to recombine, there are fewer ionising photons captured by Fe$^{{+}}$, which reduces the photoionisation rate and increases the abundance of Fe$^{{+}}$.

The significant change in the ionisation conditions in the heatboost models naturally also affects the strength of the resulting spectral lines. To illustrate this,
in the lower panels, we show the level densities (per velocity interval) of the upper levels of the [\ion{Fe}{II}] $\lambda12570$  (a$^6$D$_{9/2}\,-\,$a$^4$D$_{7/2}$) and $\lambda7155$,  (a$^4$F$_{9/2}\,-\,$a$^2$G$_{9/2}$) lines. Here, we note that the factor of four change in the loss function has altered these populations by more than an order of magnitude in much of the model and also fundamentally changed the relative importance of different regions of the model for emission of the two lines.

\subsection{Ion emission spectra}\label{sec:ionspectra}

\begin{figure*}
 \begin{center}\includegraphics[width=0.5\textwidth]{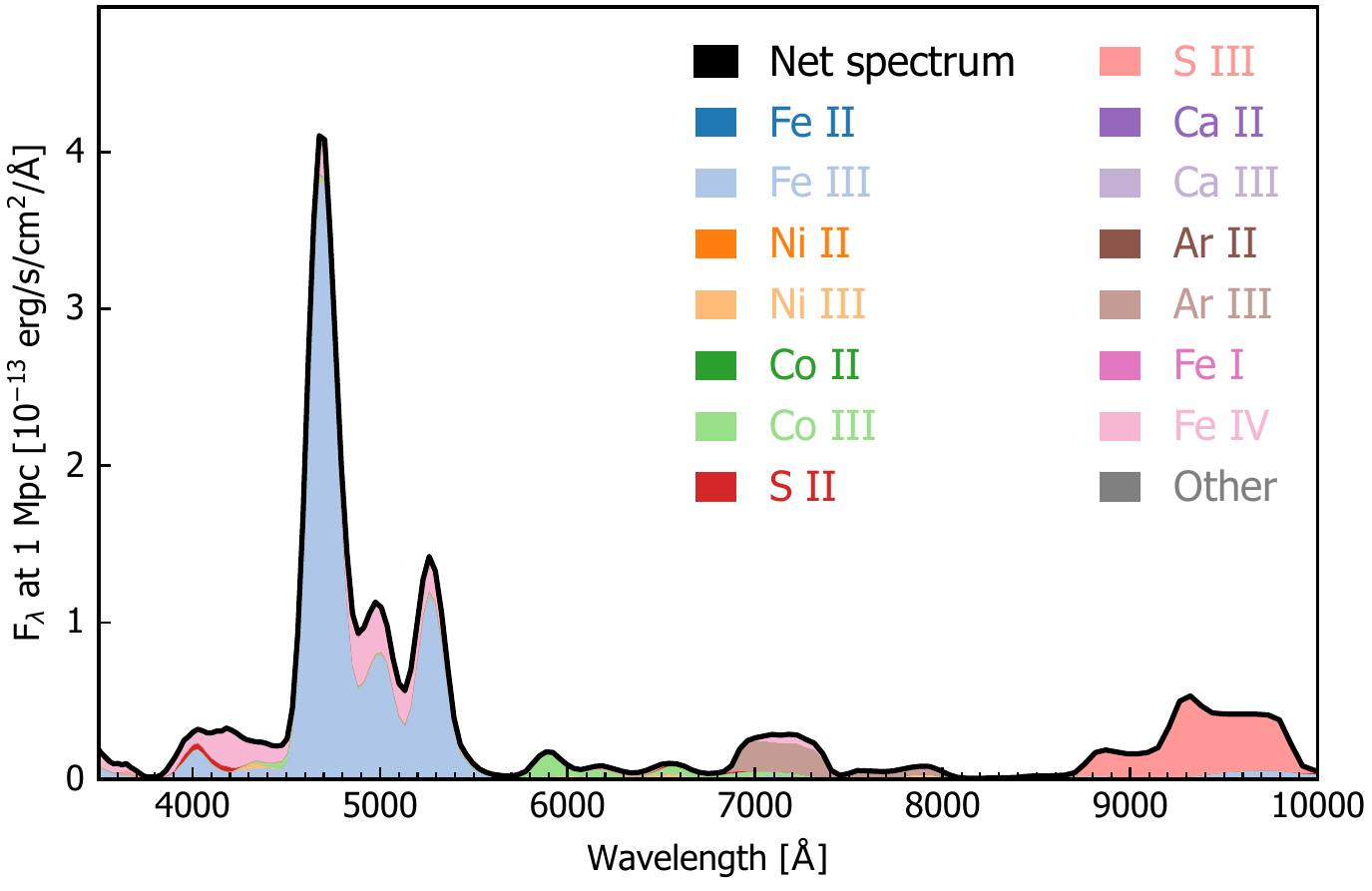}\includegraphics[width=0.5\textwidth]{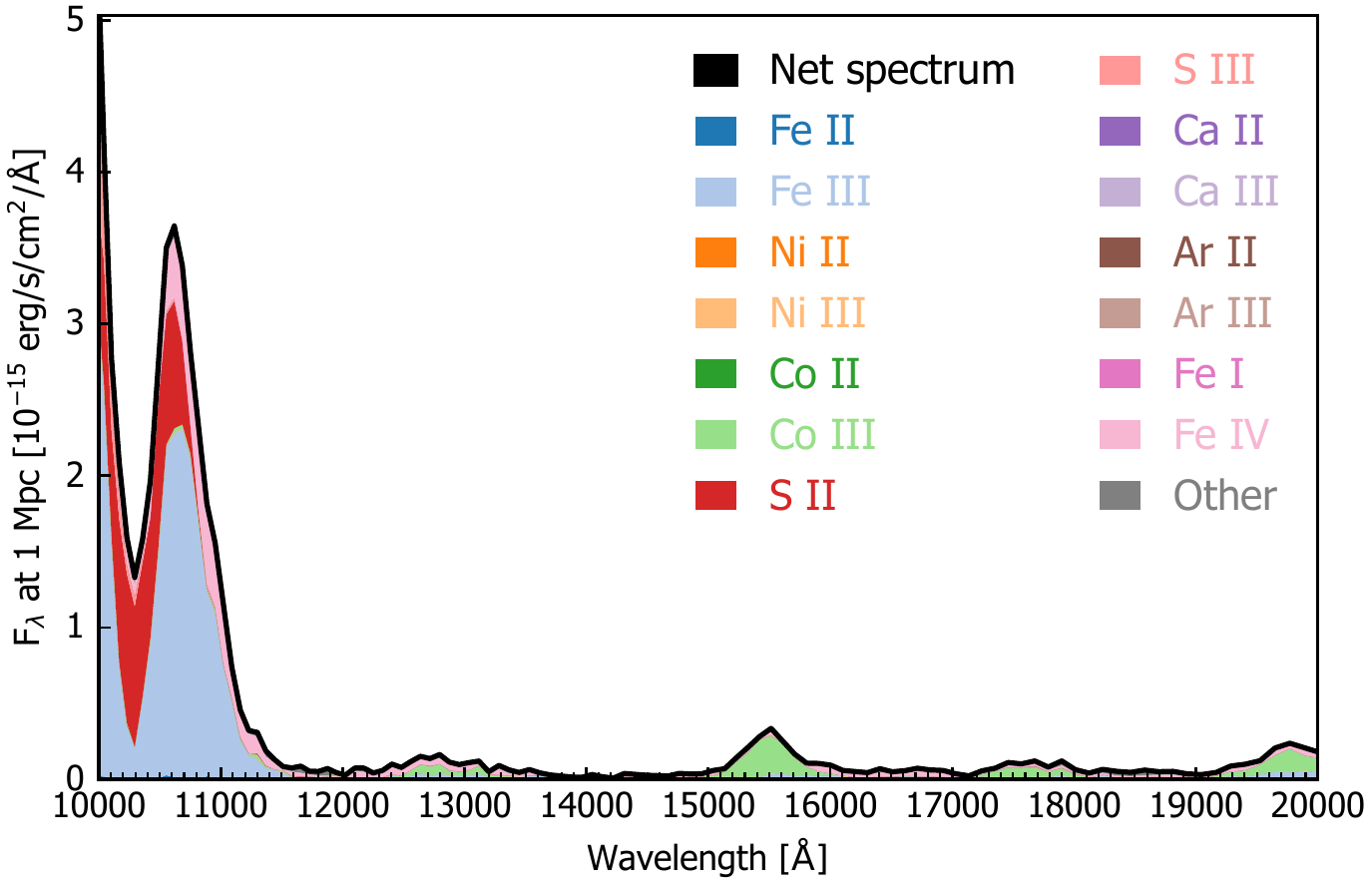}\end{center}
 \caption{Nebular emission spectra coloured by ion of the sub-\Mch model at 247 days in the optical (left) and near-infrared (right). The total spectrum is plotted as the black curve. The area under the spectrum is colour coded to indicate which ions are responsible for the emission in each wavelength bin. Note the differing scales on left and right panels.
\label{fig:emissionspectra_shen2018}}
\end{figure*}

\begin{figure*}
 \begin{center}\includegraphics[width=0.5\textwidth]{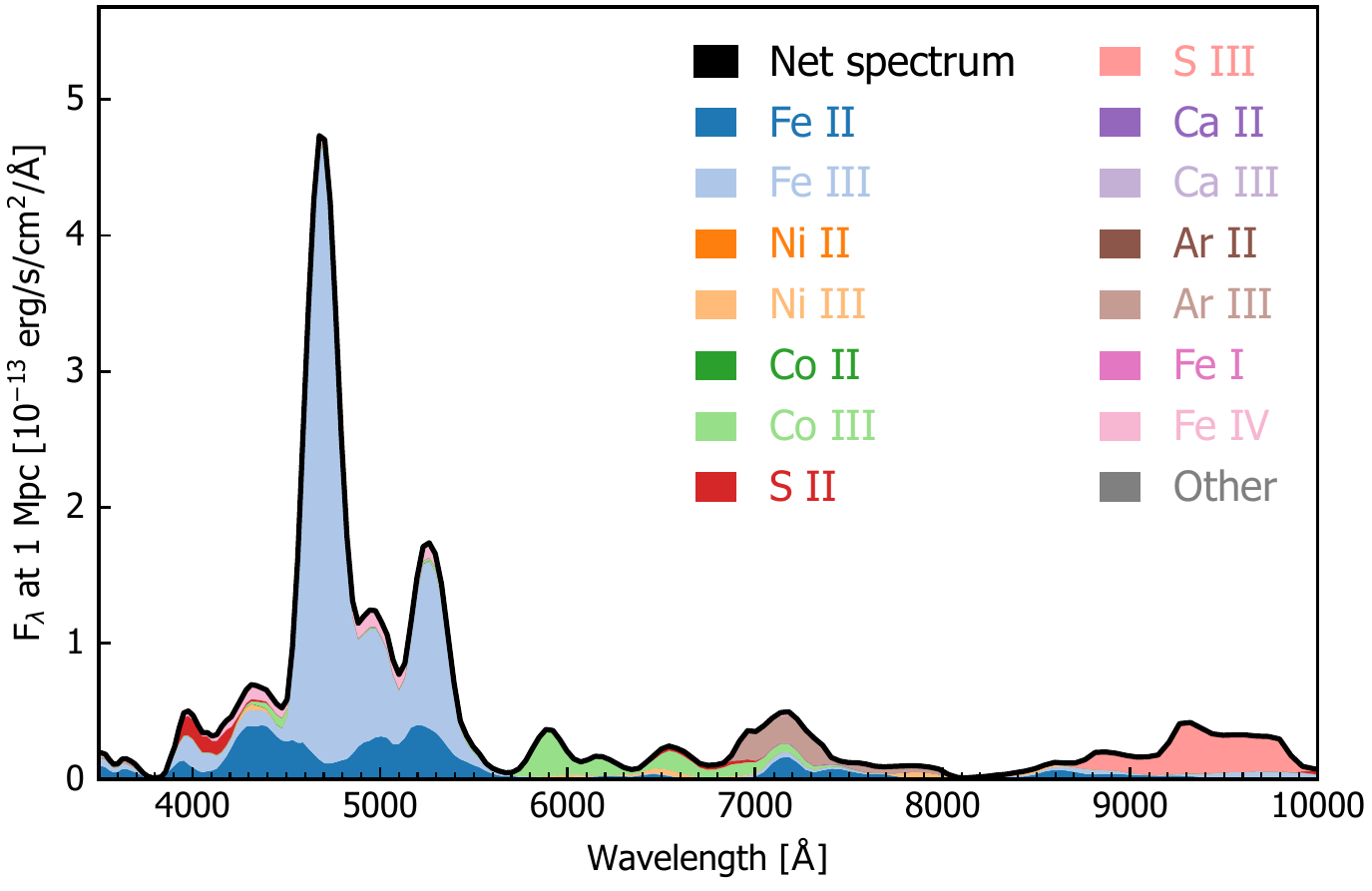}\includegraphics[width=0.5\textwidth]{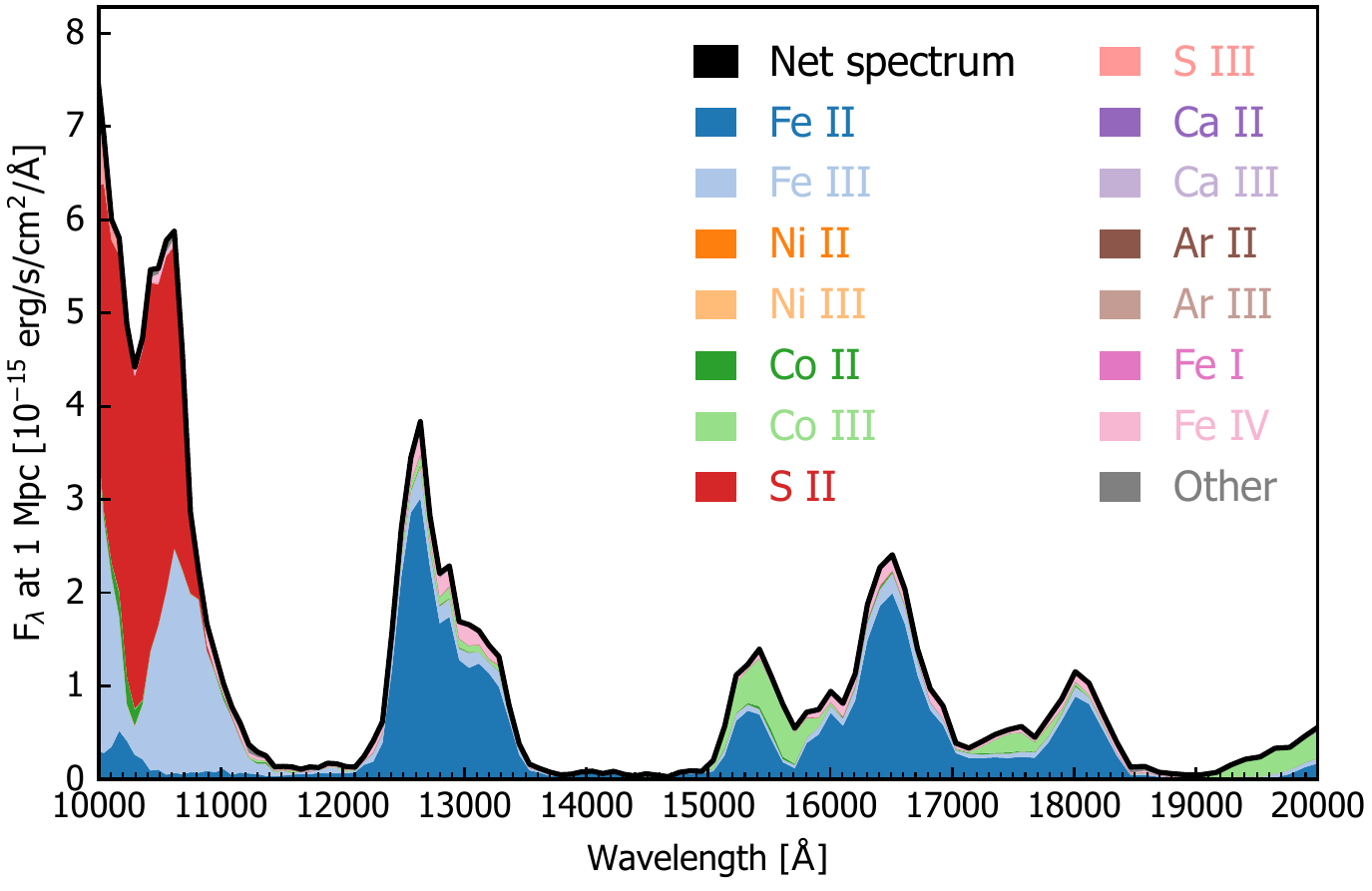}\end{center}
  \caption{Similar to Figure \ref{fig:emissionspectra_shen2018} but for the sub-\Mch \heatboostfour model at 247 days. Note the differing scales on left and right panels.
\label{fig:emissionspectra_shen2018_elossboost4x}}
\end{figure*}

\autoref{fig:emissionspectra_shen2018} and \autoref{fig:emissionspectra_shen2018_elossboost4x} show the nebular emission spectra coloured by the emitting ion for the sub-\Mch standard, and sub-\Mch \heatboostfour models, respectively.

The sub-\Mch model produces virtually no [\ion{Fe}{II}] emission features, which is most apparent in the infrared region (wavelength above 12000 \AA), where few other emission features are found in SNe Ia nebular spectra. The heatboost models increase the [\ion{Fe}{II}] line strengths to a comparable level to the W7 model \citep{Shingles:2020gy} due to the decrease in ionisation state. The increases in [\ion{Fe}{II}] line strengths are also seen in the optical region, but the difference is less clear there due to the presence of strong lines of other ions.

The [\ion{Ni}{II}] 1.939 $\upmu$m line is strong for the W7 model due to the high abundances of \iso{58}{Ni} in the core, while very little is produced in the sub-\Mch explosion model, so the reduction in the ionisation state for the heatboost model is not sufficient to produce a detectable \ion{Ni}{II} feature.

\subsection{Synthetic spectra and observations}\label{sec:spectra}

\begin{figure*}
 \begin{center}\includegraphics[width=0.5\textwidth]{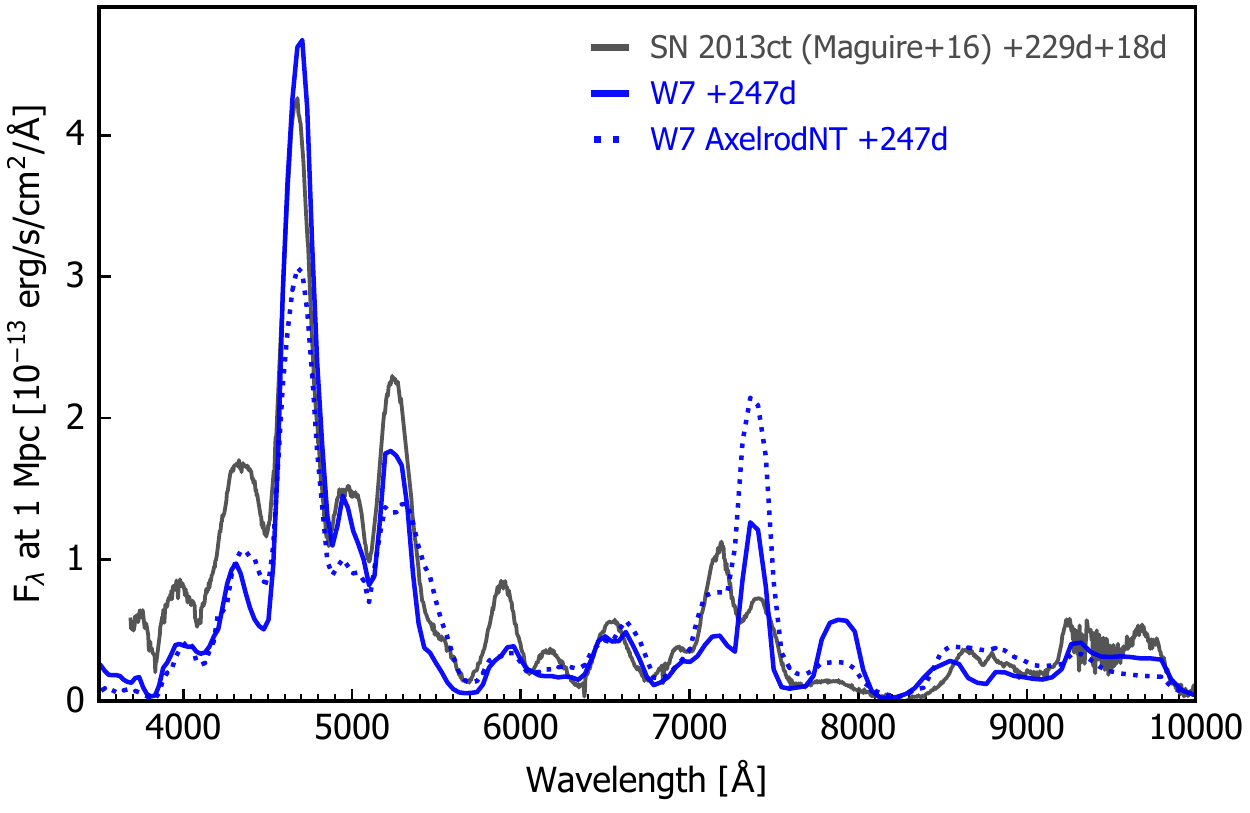}\includegraphics[width=0.5\textwidth]{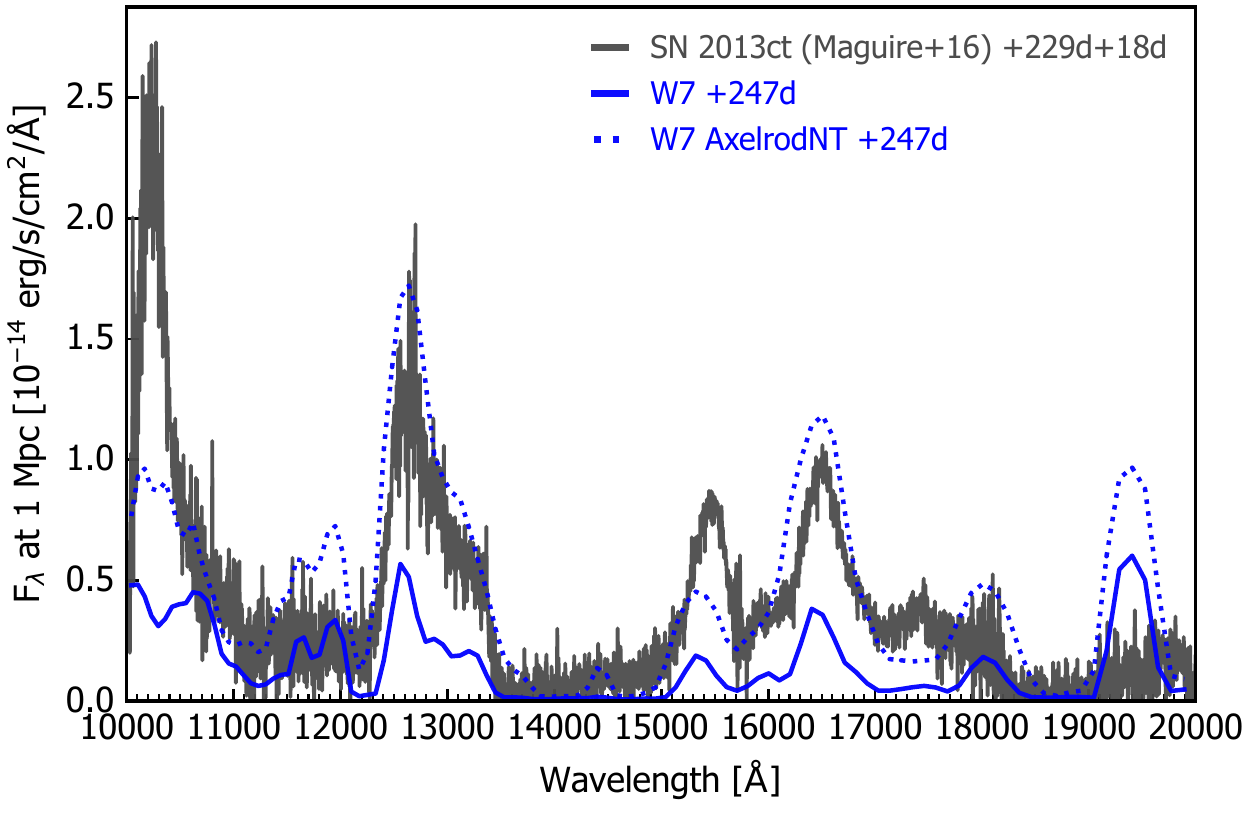}\end{center}
 \begin{center}\includegraphics[width=0.5\textwidth]{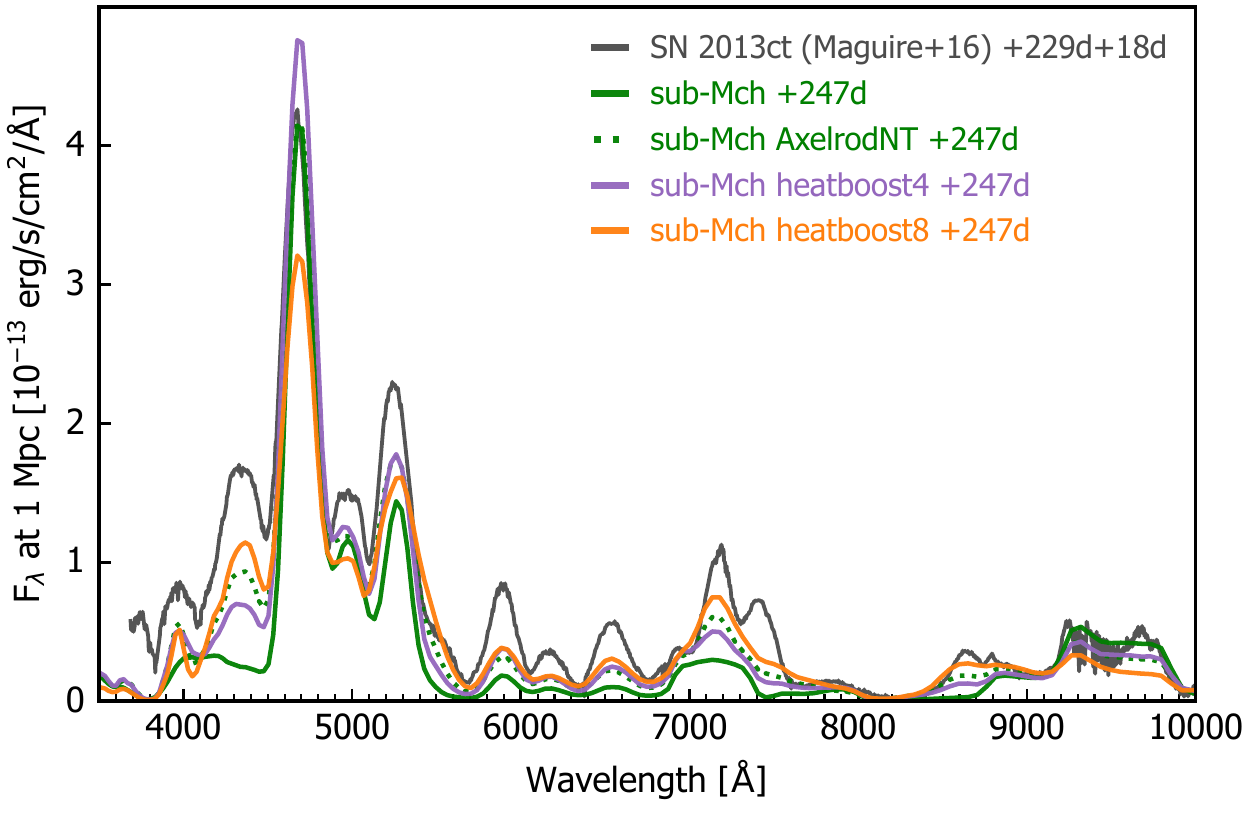}\includegraphics[width=0.5\textwidth]{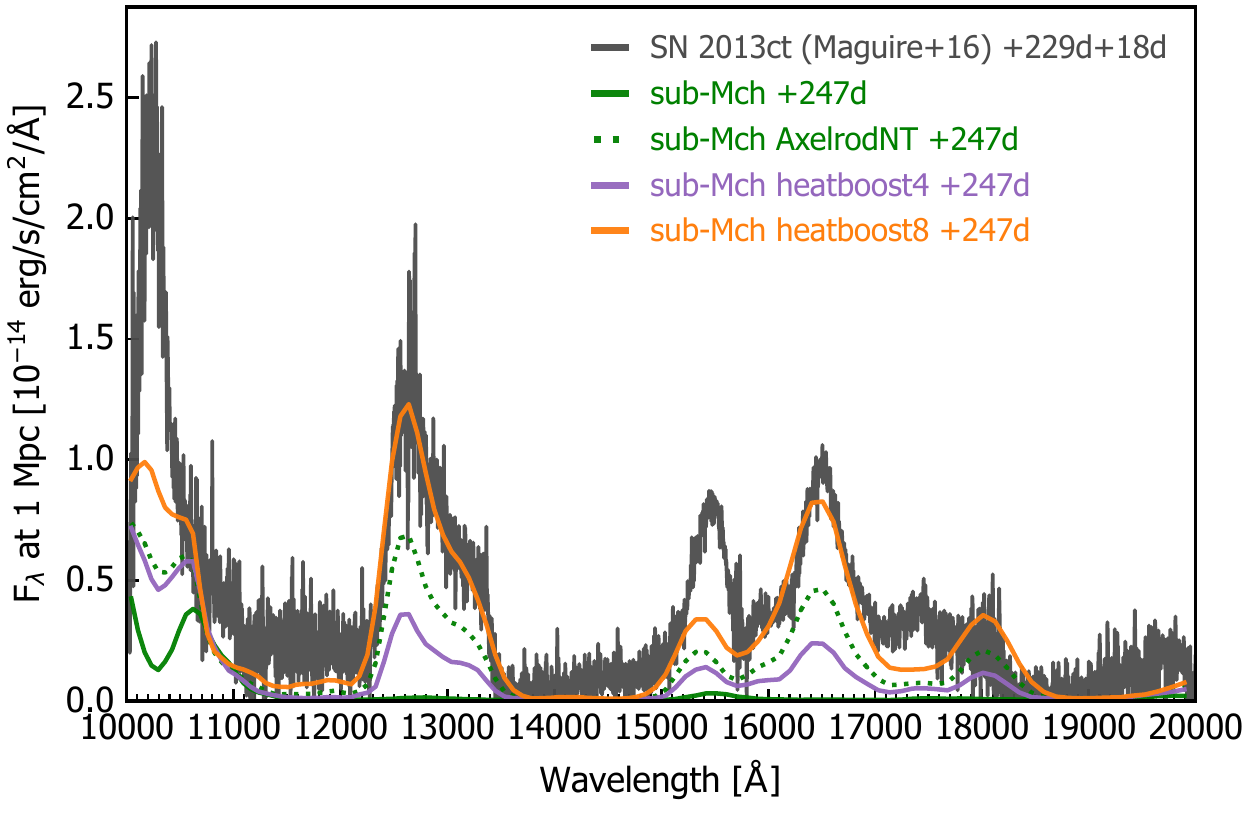}\end{center}
 \caption{Nebular spectra of W7 (top) and sub-\Mch (bottom) models at 247 days after explosion, compared to observed spectra of SN 2013ct at 229d after peak \citep{Maguire:2016jt}. A rise time of 18 days \citep{Ganeshalingam:2011ju} for the observed spectrum has been assumed.\label{fig:247d-spec}}
\end{figure*}

\autoref{fig:247d-spec} shows the optical and near-IR synthetic spectra at 247 days for the W7 \Mch \citep{Nomoto:1984jh,Iwamoto:1999jd} and sub-\Mch \citep{Shen:2018ed} models, with observed spectra from the normal Type~Ia SN 2013ct \citep{Maguire:2016jt}.

In the upper panels of \autoref{fig:247d-spec}, the W7 models with either the Spencer-Fano solution or \citet{Axelrod:1980vk} non-thermal treatments are shown with observations. As described by \citet{Shingles:2020gy}, the \artis W7 model shows significant [\ion{Fe}{II}] emission that is still weaker than SN 2013ct, a comparison most easily made in the infrared. When the AxelrodNT treatment is applied, the [\ion{Fe}{II}] lines are brought up to a similar strength as the observed spectrum, while the optical [\ion{Fe}{II}] and [\ion{Fe}{III}] features become too weak (but possibly with a more accurate ratio between the two ion stages). An anomalous feature in the W7 model around 12000\,\AA\, is made further discrepant with the observed spectrum in the AxelrodNT case, and the too-strong [\ion{Ni}{II}] emission features are made even stronger. Overall, the AxelrodNT treatment makes the fit between the W7 and SN 2013ct slightly worse.

In the lower panels  of \autoref{fig:247d-spec}, the sub-\Mch models with Spencer-Fano standard case, AxelrodNT, \heatboostfour, and \heatboosteight are compared with SN 2013ct. The most striking transition between the standard model and all other non-thermal cases is the emergence of detectable [\ion{Fe}{II}] emission.
In order of decreasing ionisation state are the \heatboostfour, AxelrodNT, and \heatboosteight models. All of these alternative cases dramatically improve the agreement with data, particularly in the infrared, as well as the optical feature around 4300\,\AA.

In summary, any of our methods to reduce the ionisation state of the sub-\Mch models dramatically improve the fit to the observed Type~Ia spectrum in a similar way.

\subsection{Line ratios and level populations}\label{sec:lineratios}

\begin{figure}
 \begin{center}\includegraphics[width=0.5\textwidth]{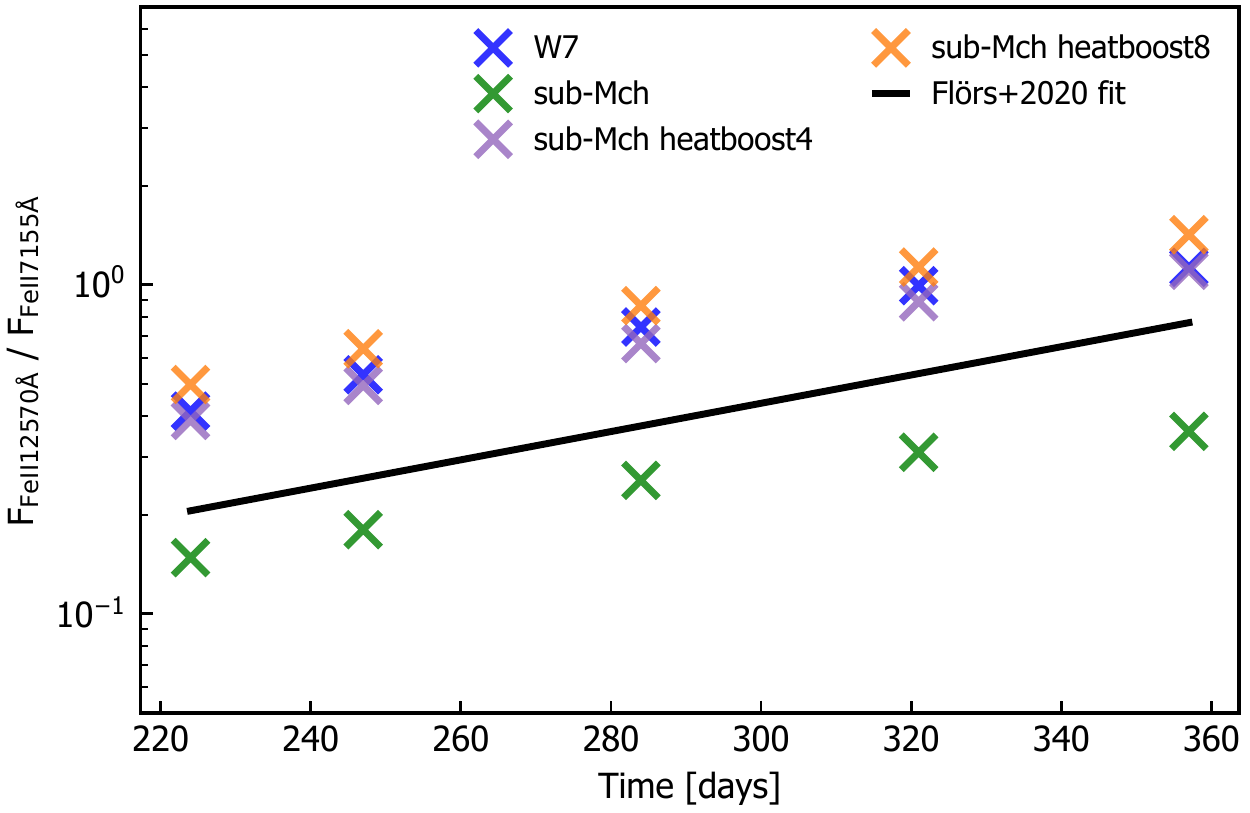}\end{center}
 \caption{The flux ratio of [\ion{Fe}{II}] $\lambda12570$ to $\lambda7155$ versus time for \artis models and the \citet{Flors:2020ek} observationally-derived line of best fit. \label{fig:fe2lineratios}}
\end{figure}

In the semi-empirical modelling approach of \citet{Flors:2020ek}, the flux ratio between the [\ion{Fe}{II}] $\lambda12570$ and $\lambda7155$ lines is measured for a sample of normal Type~Ia SNe at a range of epochs, which probes the evolution of excitation conditions over time. Here, we calculate the same line ratios from the \artis synthetic spectra of our models to test the W7 and sub-\Mch models against observations.

In \autoref{fig:fe2lineratios}, we show the [\ion{Fe}{II}] infrared-to-optical line ratios of our models, and the empirical fit relation of \citet{Flors:2020ek}. All models are close to a factor of two either above or below the empirical ratio. We find that the flux ratios are low (less IR flux) for the sub-\Mch model, while the W7, sub-\Mch \heatboostfour, and sub-\Mch \heatboosteight models are similar and above the empirical ratio. Aside from the offset, the slope of line ratio versus time is highly consistent with the empirical trend line.

\begin{figure*}
 \begin{center}\includegraphics[width=\textwidth]{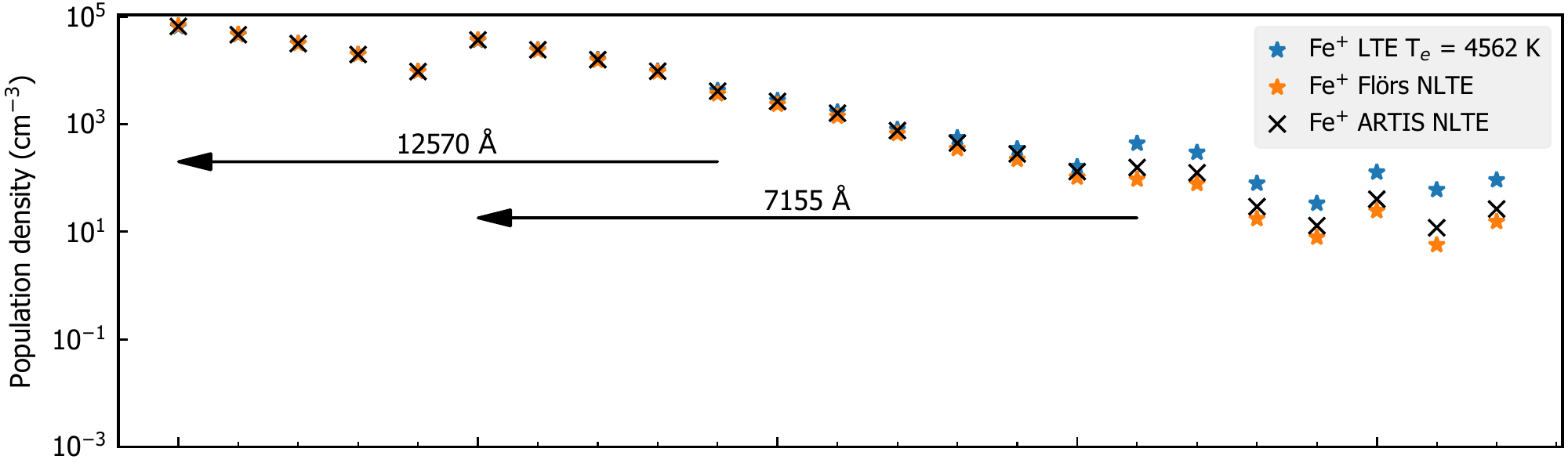}\end{center}
 \begin{center}\includegraphics[width=\textwidth]{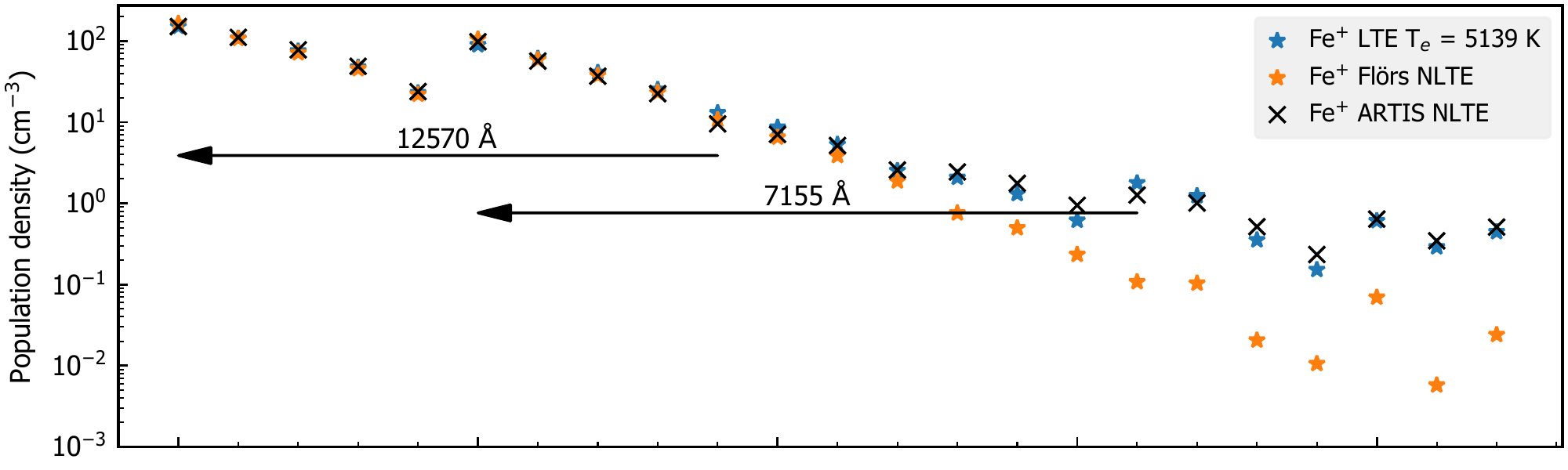}\end{center}
  \begin{center}\includegraphics[width=\textwidth]{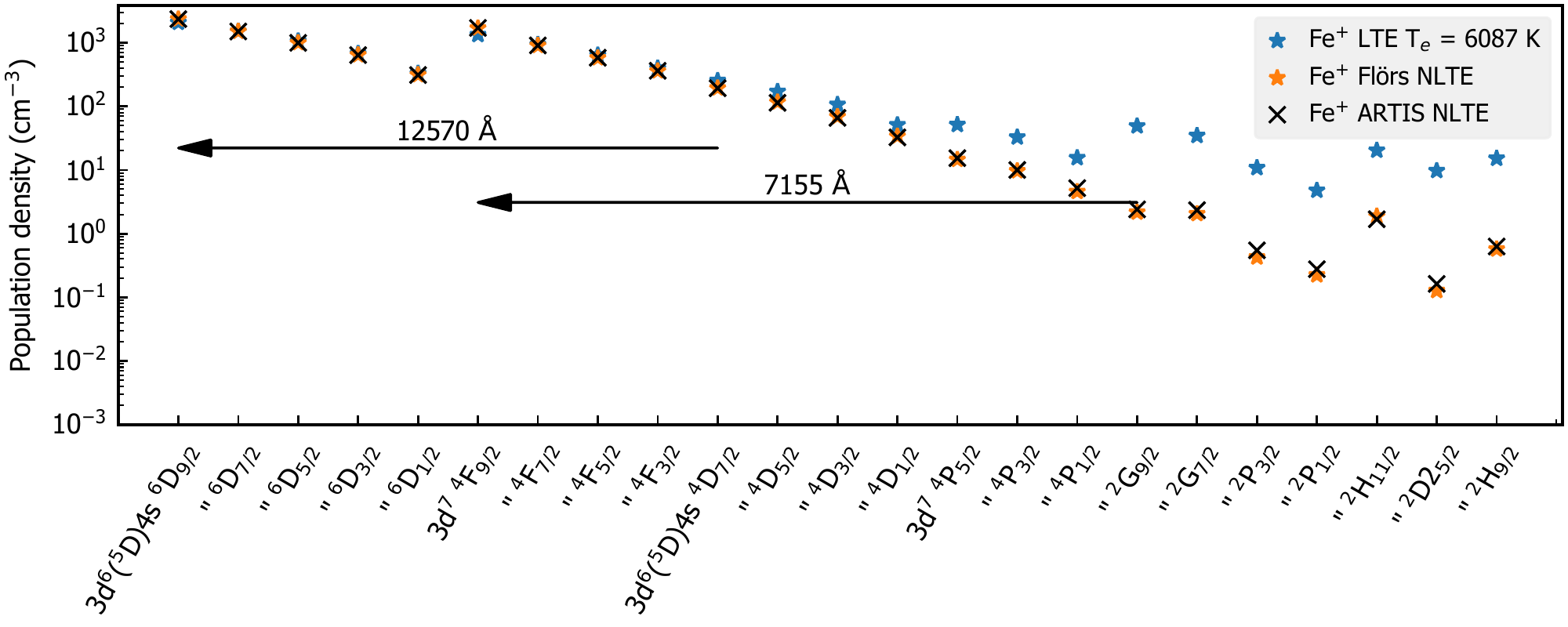}\end{center}
 \caption{The Fe$^{{+}}$ level populations at 247 days after explosion for three models near the peak radial density of Fe$^{{+}}$, and a \citet{Flors:2020ek} single-ion solution for the same respective temperature and ion population. Top: W7 4,000 km/s shell. Middle: sub-\Mch 11,000 km/s shell. Bottom: sub-\Mch \heatboosteight 11,000 km/s shell. We show the same data plotted as departure coefficients in Appendix \ref{appendix:nltedeparture}.\label{fig:nltepops_fe2}}
\end{figure*}

To explain why the line ratio of the sub-\Mch model falls below the empirical line, while the other models are very similar and above it, we look to the level populations of Fe$^{{+}}$. \autoref{fig:nltepops_fe2} shows the NLTE level populations of Fe$^{{+}}$ at 247 days in a radial zone near the peak of [\ion{Fe}{II}] emission for the W7, sub-\Mch, and sub-\Mch \heatboosteight models. We show the \artis NLTE populations, LTE populations at the same temperature, and NLTE populations from the \citet{Flors:2020ek} code (described in Section \ref{sec:parameterisedmodel}) using the \artis parameters for electron temperature, free electron density, and ion population.

For the W7 and sub-\Mch \heatboosteight models, which produce significant emission from [\ion{Fe}{II}], the level populations agree to high precision (\textless10 per cent difference) between the \artis and \citet{Flors:2020ek} calculations. This indicates that radiative excitation, photoionisation, and recombination are not important processes for setting the Fe$^{{+}}$ level population ratios in those models.

For the sub-\Mch model without heatboost, the  population of the upper level of [\ion{Fe}{II}] $\lambda7155$ which is the 16th excited state, 3d$^7$ a$^2$G$_{9/2}$, is roughly 100 times higher in the \artis model than predicted from the simplified model of \citet{Flors:2020ek}, and is close to the LTE population. When the Fe$^{{+}}$/Fe$^{2{+}}$ ratio is very small as it is in this model, the 16th excited level population becomes strongly affected by recombination cascades from Fe$^{2+}$.

In the absence of these results, we might have expected that the [\ion{Fe}{II}] line ratios would be independent of the ionisation balance, since we are considering the conditions under which [\ion{Fe}{II}] lines are emitted, rather than the overall strength of the lines, which is weak in the case of sub-\Mch models with high ionisation states. However, using the line ratios as a diagnostic of time evolution of excitation conditions requires us to have enough Fe$^{{+}}$ present that its bound-bound processes are much faster than the rate of recombination from Fe$^{2{+}}$. The presence of strong [\ion{Fe}{II}] emission in observed nebular spectra for Type~Ia SNe indicates that this regime is the one found in nature, but we are not able to factor out the ionisation state of the highly-ionised sub-\Mch model and still compare the Fe$^{{+}}$ excitation conditions to the observations. Using sub-\Mch \heatboosteight model, which has a similar temperature structure to the model with no heatboost, is one way to compare the sub-\Mch model to the observations by bringing the ionisation state toward a more realistic level and avoiding the recombination-cascade effects on the line ratios. However, neither the W7 and sub-\Mch \heatboosteight model is favoured on the basis of their match to empirical infrared-to-optical line ratios because they are too high by a very similar factor (\autoref{fig:fe2lineratios}). This shows that the semi-empirical method to understand the excitation state of Fe is not a strong discriminant between the different explosion models that do produce [\ion{Fe}{II}] emission features.

\subsection{Line-emitting conditions and one-zone approximation}\label{sec:emittingconditions}
In the analysis of observed spectra, most studies adopt a one-zone approximation, assuming that the line emissions are well-characterised by some value for the electron density and the electron temperature. In contrast, sophisticated radiative transfer models form emission lines from the contributions of multiple zones. Here we test the validity of the one-zone assumption by measuring the range of conditions under which the [\ion{Fe}{II}] lines are formed in our models.

\begin{figure*}
 \begin{center}
	 \includegraphics[width=0.45\textwidth]{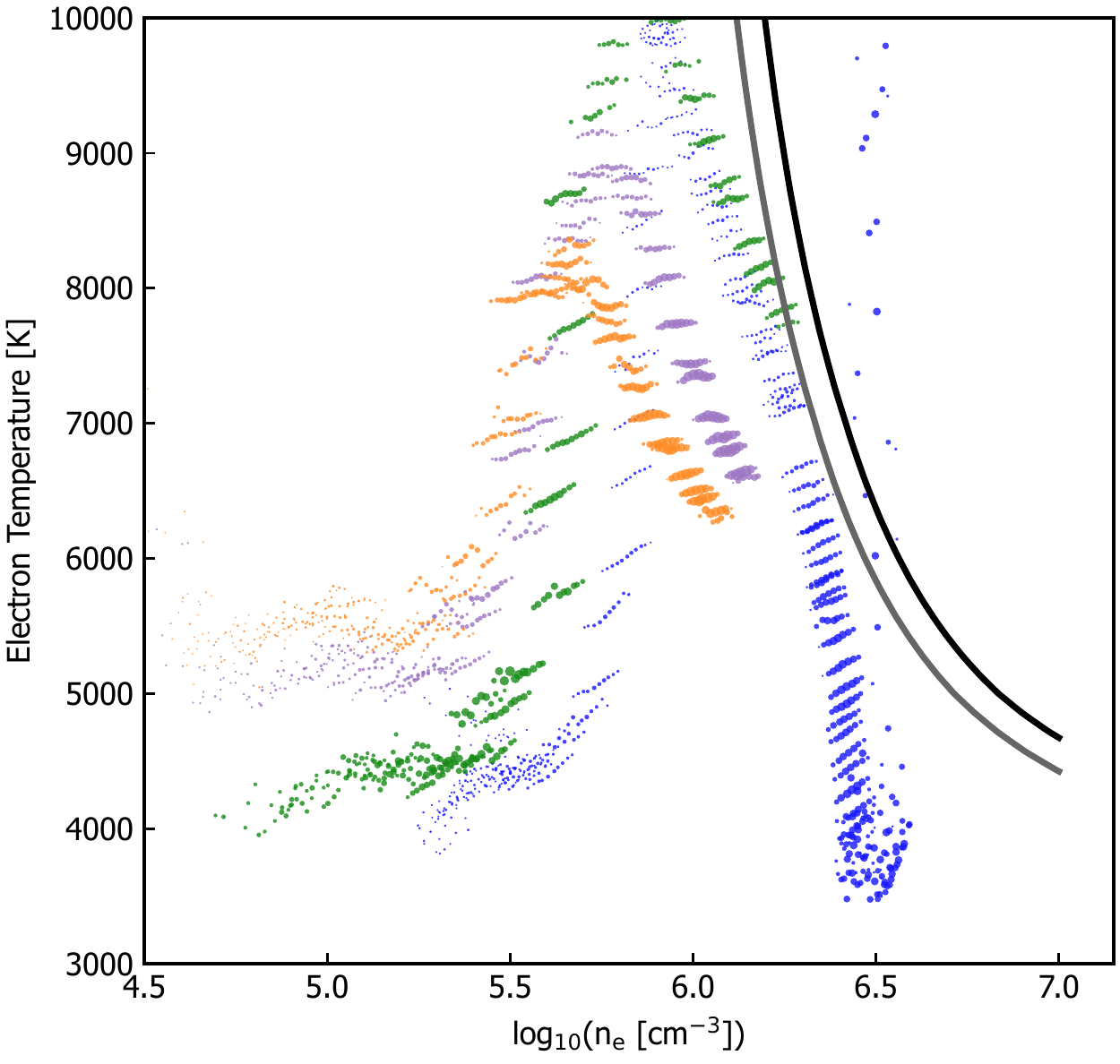}
	 \includegraphics[width=0.45\textwidth]{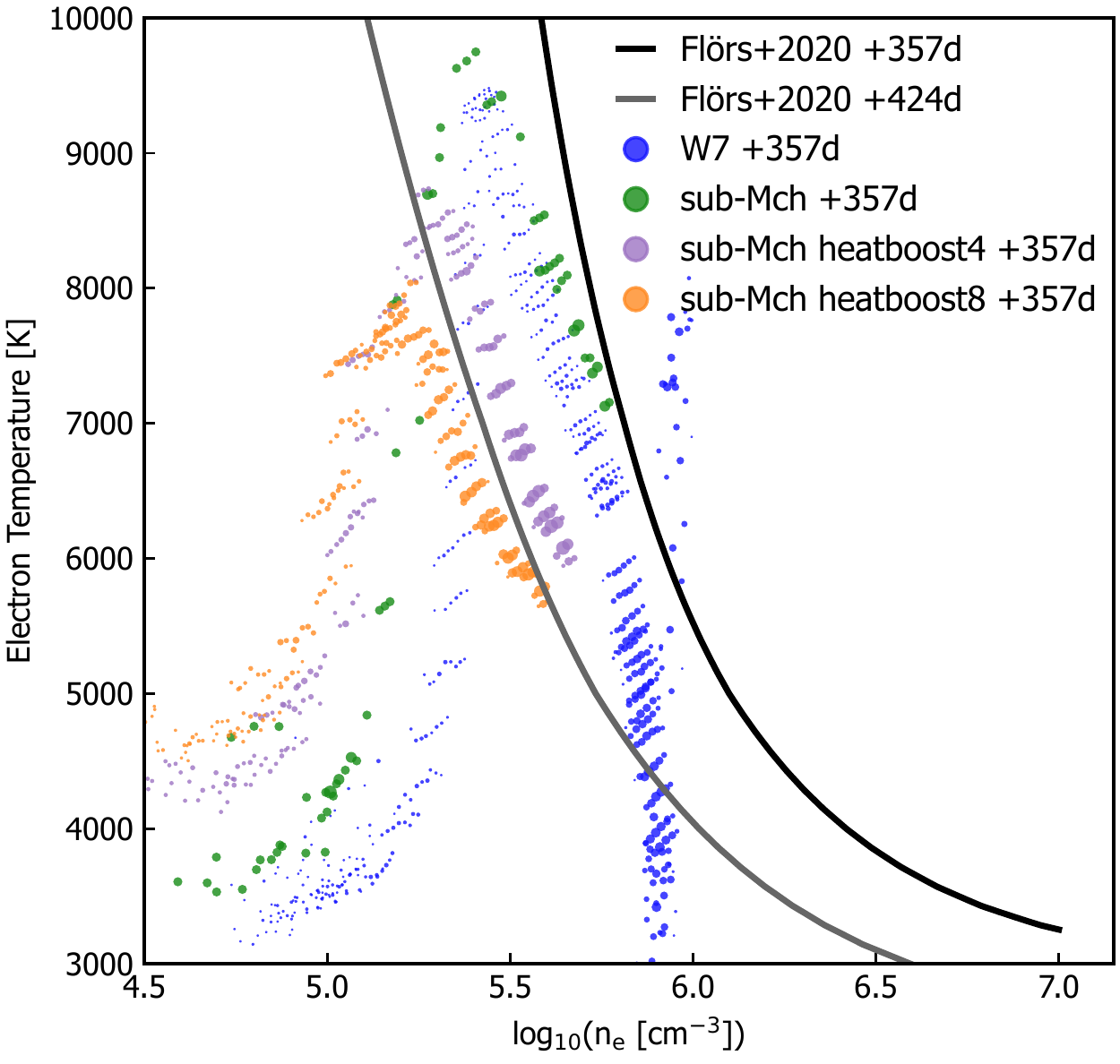}
 \end{center}
 \caption{Line emission conditions for [\ion{Fe}{II}] $\lambda12570$ and $\lambda7155$ lines. Each circle (area proportional to flux, normalised to each model's flux in the two lines) represents a shell from which [\ion{Fe}{II}] lines are emitted in electron temperature versus free electron density plane. The tracks of points represent single shells over multiple timesteps within the range of $\pm 5$ days of the specified time. Also shown are the empirically-derived best-fit lines from \citet{Flors:2020ek}.
 \label{fig:telecdens}}
\end{figure*}

In Figure \ref{fig:telecdens}, we plot the model shells as points in the temperature versus electron density plane with an area proportional to the outgoing flux in the [\ion{Fe}{II}] $\lambda12570$ and $\lambda7155$ lines. For comparison to the models, Figure \ref{fig:telecdens} also shows the empirically-derived \citet{Flors:2020ek} curve that would be required for a one-zone, collisional-radiative NLTE model to match the [\ion{Fe}{II}] infrared-to-optical line ratio. There is a degeneracy between the two parameters, because they both scale up or down the rates of collisional transitions. It can be seen by comparison to the \citet{Flors:2020ek} curves that the line-forming regions of our models are systematically either lower temperatures, lower densities, or both.

Even though Figure \ref{fig:fe2lineratios} suggests that the physical conditions in $n_e$-T space can lie on either side of the empirical relation (see black line in Fig.~\ref{fig:telecdens}), all of the models we investigated exhibit temperatures and electron densities below this curve. While in the case of the W7 and the heatboost sub-\Mch model, the temperature and free electron density are lower than what is expected from the analysis of observed SNe\,Ia. The result for the standard sub-\Mch model is a direct consequence of the fact that the Fe$^{{+}}$ upper level of [\ion{Fe}{II}] $\lambda7155$ is boosted by recombination of Fe$^{2+}$, thus increasing the strength of this transition relative to the unaffected $\lambda12570$ line.

\section{Discussion and Conclusions}
We have investigated the discrepancy between conclusions reached about the progenitors of Type~Ia SNe from semi-empirical analyses of Type~Ia nebular spectra, and forward modelling of theoretical explosion simulations.

We have showed that the ratios of [\ion{Fe}{II}] lines are not always completely independent from the overall flux of [\ion{Fe}{II}] and the ionisation balance. Specifically, when the Fe$^{{+}}$/Fe$^{2{+}}$ ratio is very small, the level populations of Fe$^{{+}}$ become strongly affected by the recombination cascades. This fact prevents the line ratios of singly-ionised species from being used as an independent diagnostic of explosion models that are too highly ionised. Therefore, meaningfully constraining the progenitor problem of Type~Ia SNe will require us to understand how to accurately model the ionisation state under the relevant ejecta conditions.

In agreement with \citet{Wilk:2020kb}, we have shown that the ionisation state is a major factor that determines the
overall goodness-of-fit between synthetic nebular spectra from hydrodynamical explosion models and the observed data. Using what is widely-considered to be the most accurate treatment of non-thermal ionisation, solving the Spencer-Fano equation with the best-available cross sections, leads to the conclusion that sub-\Mch models are too highly ionised to reproduce the [\ion{Fe}{II}]-dominated spectra of observed Type~Ia SNe at late times. Reducing the ionisation state by reducing the non-thermal ionisation rates leads to better agreement with the data for the sub-\Mch model, but no physical justification for reducing these rates has yet been found.

Our parameter exploration is not a physical explanation for the modelled ionisation state being too high. Our method is to alter one numerical parameter with the effect of decreasing the non-thermal ionisation rate by directing the non-thermal energy toward heating. A candidate physical process that could also have the effect of reducing the ionisation state is clumping, as discussed in the context of Type~Ia SNe by \citet{Wilk:2020kb}. Clumping leads to increased recombination rates, which reduces the ionisation state. However, increased clumping can have secondary effects, such as causing anomalously high emission from [\ion{Ca}{II}] in near-\Mch models that assume a constant clumping factor throughout the ejecta \citep{Wilk:2020kb}. It is unclear if reducing the non-thermal ionisation rates would avoid such anomalous features. It is plausible that reducing the non-thermal ionisation rates might lead to different spectra compared to effectively boosting the recombination rates, such as by clumping. This is because the ionisation rates of some ions are dominated by photoionisation, so they would be only weakly affected by changes to the non-thermal ionisation rate, while changes to recombination rate affect all ions. There are also consequences for the photoionisation rates that are difficult to predict in advance, because they are linked with the recombinations of other ions.

One strategy to disentangle reduced non-thermal ionisation rates from clumping would be to make a time-series analysis of ionisation state in comparison to observed SNe Ia. This is because the influence of non-thermal ionisation on the ionisation state only becomes important around several tens of days after explosion, as the rates of photoionisation decline with the expanding and cooling ejecta. In contrast, increasing the recombination rates, such as by clumping, would decrease the ionisation state at all times, including the very early times.

Further research into the effects of clumping and possible improvements to the non-thermal ionisation accuracy will be needed to see if a single model can accurately reproduce the ionisation state of all elements that affect the nebular spectra of Type~Ia supernovae. This work will also contribute to advancements in kilonova modelling.

\section*{Acknowledgements}
The authors thank D. John Hillier for making the \cmfgen atomic data set publicly available.

SS and LS acknowledge support from STFC through grant, ST/P000312/1.
LS and AF acknowledge support by the European Research Council (ERC) under the European Union’s Horizon 2020 research and innovation program (ERC Advanced Grant KILONOVA No. 885281). CC acknowledges support by the European Research Council (ERC) under the European Union’s Horizon
2020 research and innovation program under grant agreement No. 759253. KJS is supported by NASA through the Astrophysics Theory Program (NNX17AG28G).

The work was supported through computational resources at Forschungszentrum J{\"u}lich via project HMU14.

This research was undertaken with the assistance of resources, in particular the UNSW Merit Allocation Scheme and the Flagship Allocation Scheme, from the National Computational Infrastructure (NCI), which is supported by the Australian Government.

This work was performed using the Cambridge Service for Data Driven Discovery (CSD3), part of which is operated by the University of Cambridge Research Computing on behalf of the STFC DiRAC HPC Facility (www.dirac.ac.uk). The DiRAC component of CSD3 was funded by BEIS capital funding via STFC capital grants ST/P002307/1 and ST/R002452/1 and STFC operations grant ST/R00689X/1. DiRAC is part of the National e-Infrastructure.

This research has made use of NASA's Astrophysics Data System.
Figures in this work have been generated with the \textsc{matplotlib} package \citep{Hunter:2007ih} and \textsc{artistools}\footnote{\href{https://github.com/artis-mcrt/artistools/}{https://github.com/artis-mcrt/artistools/}}.

\section*{Data availability}
The data underlying this article will be shared on reasonable request to the corresponding author.

\bibliographystyle{mnras}
\bibliography{references}

\appendix
\section{NLTE populations}\label{appendix:nlte}
\begin{figure*}
 \begin{center}\includegraphics[width=\textwidth]{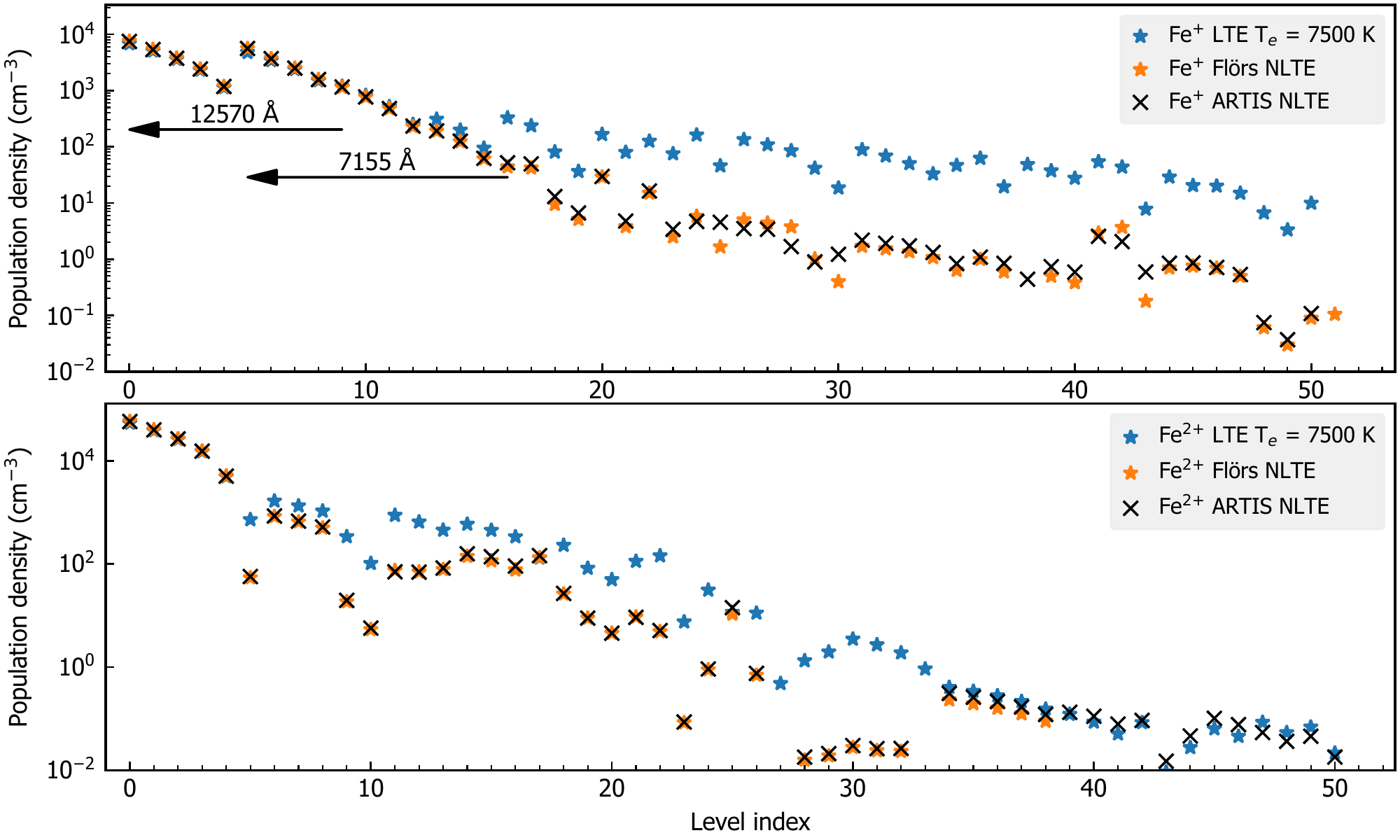}\end{center}
 \caption{NLTE populations for Fe$^{{+}}$ and Fe$^{{2+}}$ calculated with \artis and the Flörs code for a collisional-radiative model at the same temperature, electron density, and Fe density. \label{fig:nltetest}}
\end{figure*}

A NLTE population solver comparison for \ion{Fe}{II} and \ion{Fe}{III} is shown in Figure \ref{fig:nltetest}. NLTE points are the ARTIS calculations, which are shown to be identical to Flörs NLTE solution for Fe$^{2{+}}$ and near-identical for Fe$^{{+}}$. We include collisional excitation/de-excitation and radiative decays. The main point of difference is the presence of recombination transitions in ARTIS, as this method simultaneously solves the level populations for all included ionisation stages of the element. With similar populations in both ionisation stages, the recombination process does not significantly affect the level populations.

\section{Departure ratio plots}\label{appendix:nltedeparture}
\begin{figure*}
 \begin{center}\includegraphics[width=\textwidth]{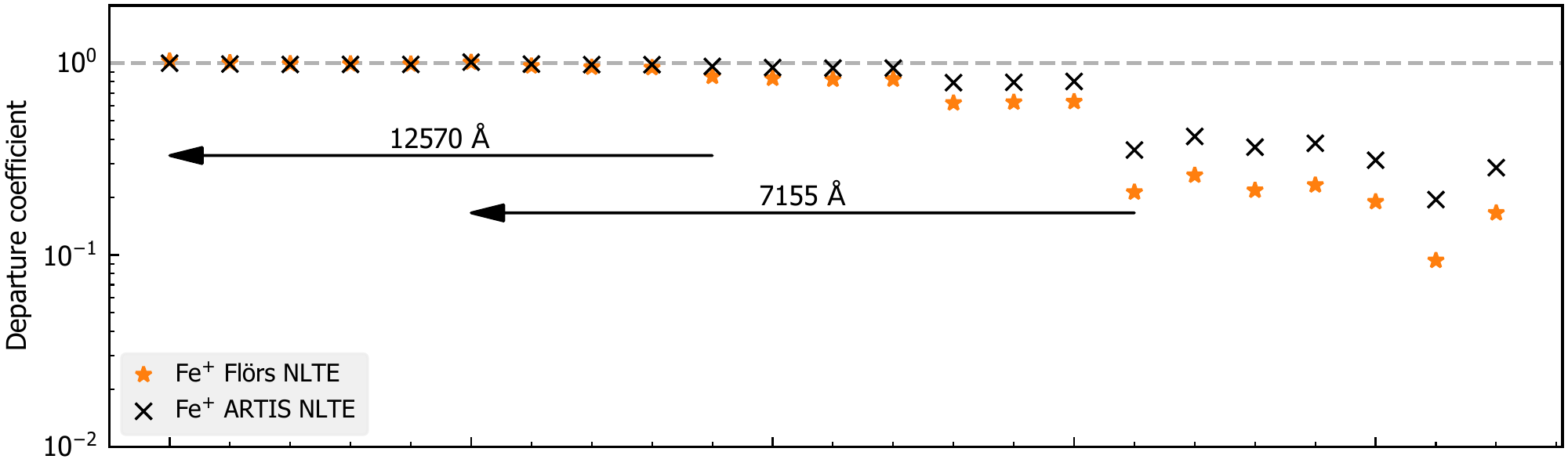}\end{center}
 \begin{center}\includegraphics[width=\textwidth]{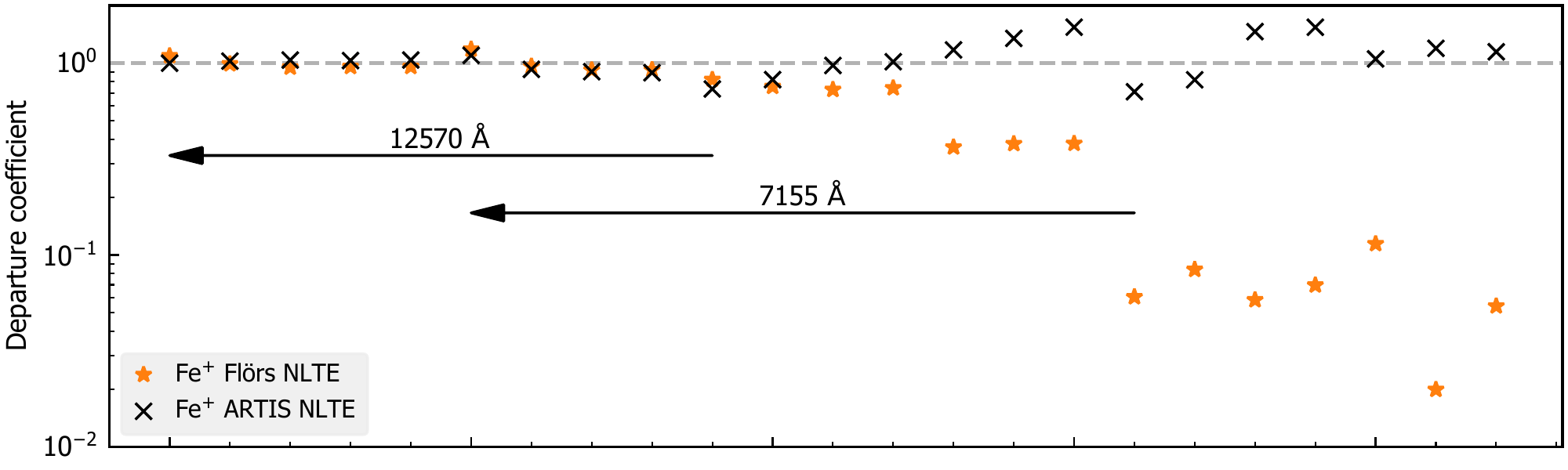}\end{center}
  \begin{center}\includegraphics[width=\textwidth]{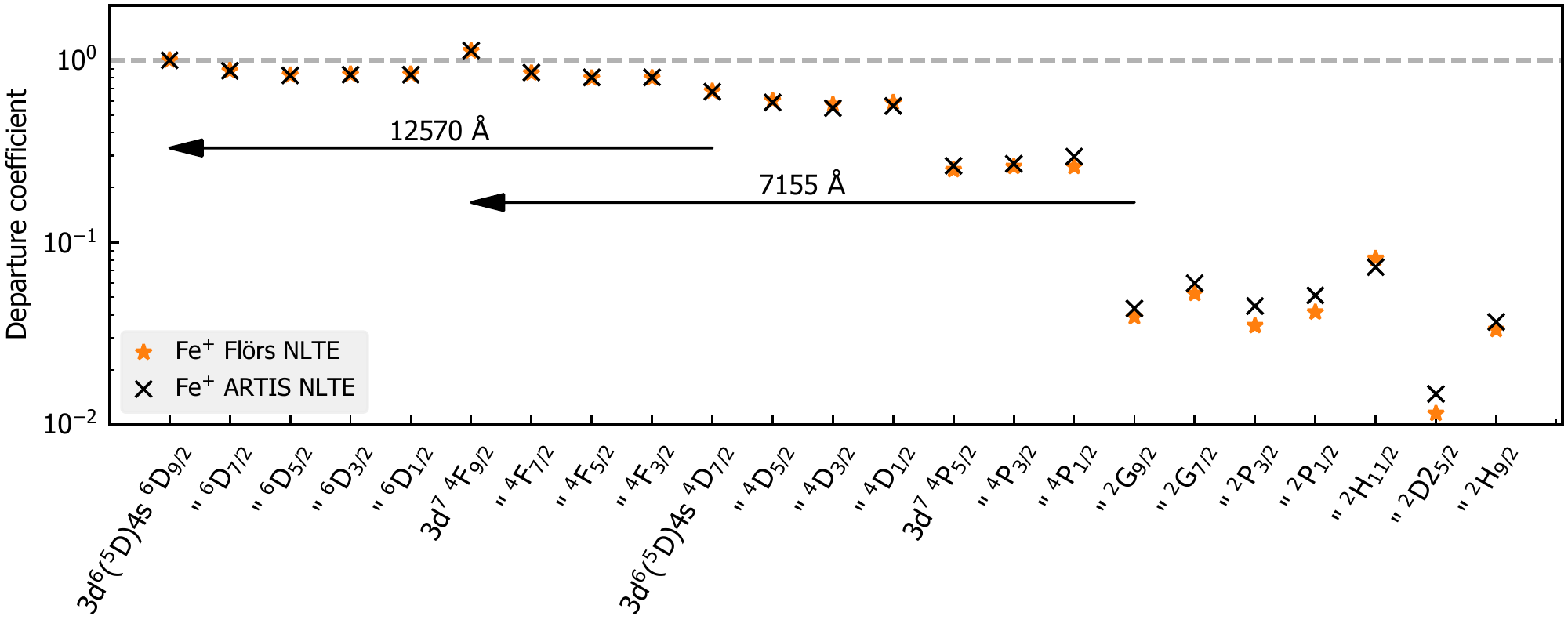}\end{center}
 \caption{The Fe$^{{+}}$ level departure coefficients at 247 days after explosion for three models near the peak radial density of Fe$^{{+}}$, and the \citet{Flors:2020ek} single-ion calculation for the same respective temperature and ion population. Top: W7 4,000 km/s shell. Middle: sub-\Mch 11,000 km/s shell. Bottom: sub-\Mch \heatboosteight 11,000 km/s shell.\label{fig:nltepops_fe2_departure}}
\end{figure*}

In Figure \ref{fig:nltepops_fe2_departure}, we present the same data as Figure \ref{fig:nltepops_fe2}, except that each level population has been divided by its population under LTE conditions at the same local electron temperature and ion population.

\section{Bound electron loss rate}\label{appendix:boundlossrate}
Here, we compare the experimentally-calibrated stopping power for high-energy electrons in a medium to the energy losses calculated from by our set of ionisation and excitation transitions. For a quantitative comparison, we use the example of a 100 keV electron in a neutral Fe gas with a number density of $10^5$ cm$^{-3}$.

Using \citet{Barnes:2016fu} equation 5 for the total loss rate \citep{Heitler:1954uu,Berger:1964vc,Chan:1993ic,Milne:1999gw},
\begin{align*}
    \frac{dE}{dt} &= \frac{2 \pi r_0^2 m_e c^3 n_{e,b}}{\beta} \left[2 \ln\left(\frac{E}{I}\right) + \ln\left(1 + \frac{\tau}{2}\right) + \left(1 - \frac{v^2}{c^2} \right) g(\tau)\right]\\
    &\text{where } g(\tau)=1 + \frac{\tau^2}{8} - (2\tau + 1)\ln 2\\
        &= 0.424 \,\mathrm{eV/s}
\end{align*}
where $\tau=E/m_ec^2$, and $I = 287.8$ eV as given by equation 7 of \citet{Barnes:2016fu}. The NIST ESTAR database \citep{Estar:2017iu} reports the stopping power in Fe of 2.83 MeV cm$^2$/g, which when converted to a loss rate at our example density (0.434 eV/s), agrees very closely with the formula used by Barnes.



Using excitation and impact ionisation cross sections for neutral Fe and the average energy loss of the primary \citep[with the same cross sections and secondary energy distribution function as][]{Kozma:1992cy}, the atomic loss rate is
\begin{align*}
    \frac{dE}{dt} &= \left[ \sum_{e} \sigma_e(E) \cdot \epsilon_e(E) + \sum_i \sigma_i(E) \cdot \left<\epsilon_i(E)\right> \right] \cdot n_i \cdot v\\
        &= 0.188 \,\mathrm{eV/s}.
\end{align*}
This total is clearly too low compared to the total stopping power, suggesting some combination of missing physical processes or an underestimate of energy loss of the primary electron in each interaction.

For reference, the plasma loss rate \citep{Schunk:1971fs} for a 100 keV electron to the thermal free electrons (with number density $10^5$ cm$^{-3}$) is:
\begin{align*}
    \frac{dE}{dt} &= n_e v \frac{2\pi e^4}{E} \ln  \left(\frac{2E}{\zeta_e}\right)\mathrm{, where \, }\zeta_e = \frac{h}{2\pi}\omega_p = \frac{h}{2\pi} \sqrt{\frac{4\pi n_e e^2}{m_e}}\\
        &= 0.065 \,\mathrm{eV/s}
\end{align*}

\label{lastpage}
\end{document}